\documentclass[12pt]{article}

\pdfoutput=1

\usepackage{graphicx}
\usepackage{dcolumn}
\usepackage{bm}
\usepackage{hyperref}
\usepackage{color}
\usepackage{amsmath}
\usepackage{amssymb}
\usepackage{amsfonts}
\usepackage{multirow}
\usepackage{cite}

\usepackage[top=30truemm,bottom=30truemm,left=25truemm,right=25truemm]{geometry}

\begin{document}


\def\a{\alpha}
\def\b{\beta}
\def\c{\varepsilon}
\def\d{\delta}
\def\e{\epsilon}
\def\f{\phi}
\def\g{\gamma}
\def\h{\theta}
\def\k{\kappa}
\def\l{\lambda}
\def\m{\mu}
\def\n{\nu}
\def\p{\psi}
\def\q{\partial}
\def\r{\rho}
\def\s{\sigma}
\def\t{\tau}
\def\u{\upsilon}
\def\v{\varphi}
\def\w{\omega}
\def\x{\xi}
\def\y{\eta}
\def\z{\zeta}
\def\D{\Delta}
\def\G{\Gamma}
\def\H{\Theta}
\def\L{\Lambda}
\def\F{\Phi}
\def\P{\Psi}
\def\S{\Sigma}

\def\o{\over}
\def\beq{\begin{align}}
\def\eeq{\end{align}}
\newcommand{\gsim}{ \mathop{}_{\textstyle \sim}^{\textstyle >} }
\newcommand{\lsim}{ \mathop{}_{\textstyle \sim}^{\textstyle <} }
\newcommand{\vev}[1]{ \left\langle {#1} \right\rangle }
\newcommand{\bra}[1]{ \langle {#1} | }
\newcommand{\ket}[1]{ | {#1} \rangle }
\newcommand{\EV}{ {\rm eV} }
\newcommand{\KEV}{ {\rm keV} }
\newcommand{\MEV}{ {\rm MeV} }
\newcommand{\GEV}{ {\rm GeV} }
\newcommand{\TEV}{ {\rm TeV} }
\newcommand{\1}{\mbox{1}\hspace{-0.25em}\mbox{l}}
\newcommand{\headline}[1]{\noindent{\bf #1}}
\def\diag{\mathop{\rm diag}\nolimits}
\def\Spin{\mathop{\rm Spin}}
\def\SO{\mathop{\rm SO}}
\def\O{\mathop{\rm O}}
\def\SU{\mathop{\rm SU}}
\def\U{\mathop{\rm U}}
\def\Sp{\mathop{\rm Sp}}
\def\SL{\mathop{\rm SL}}
\def\tr{\mathop{\rm tr}}
\def\mpl{M_{\rm Pl}}

\def\IJMP{Int.~J.~Mod.~Phys. }
\def\MPL{Mod.~Phys.~Lett. }
\def\NP{Nucl.~Phys. }
\def\PL{Phys.~Lett. }
\def\PR{Phys.~Rev. }
\def\PRL{Phys.~Rev.~Lett. }
\def\PTP{Prog.~Theor.~Phys. }
\def\ZP{Z.~Phys. }

\def\dd{\mathrm{d}}
\def\ff{\mathrm{f}}
\def\BH{{\rm BH}}
\def\inf{{\rm inf}}
\def\ev{{\rm evap}}
\def\eq{{\rm eq}}
\def\SM{{\rm sm}}
\def\Mpl{M_{\rm Pl}}
\def\GeV{{\rm GeV}}

\def\Msusy{m_{\rm stop}}
\newcommand{\Red}[1]{\textcolor{red}{#1}}
\newcommand{\TL}[1]{\textcolor{blue}{\bf TL: #1}}

\baselineskip 0.7cm
\begin{titlepage}

\vskip 1.35cm
\begin{center}

{\bf\large Minimal Non-Abelian Supersymmetric Twin Higgs}

\vskip 1.2cm
Marcin Badziak$^{1,2,3}$ and Keisuke Harigaya$^{2,3}$
\vskip 0.4cm
$^1${\it Institute of Theoretical Physics, Faculty of Physics, University of Warsaw, ul.~Pasteura 5, PL--02--093 Warsaw, Poland}\\
$^2${\it Department of Physics, University of California, Berkeley, California 94720, USA}\\
$^3${\it Theoretical Physics Group,  Lawrence Berkeley National Laboratory, Berkeley, California 94720, USA}
\vskip 1.5cm

\abstract{
We propose a minimal supersymmetric Twin Higgs model that can accommodate tuning of the electroweak scale for heavy stops better than $10\%$ with high mediation scales of supersymmetry breaking. A crucial ingredient of this model is a new $SU(2)_X$ gauge symmetry which provides a $D$-term potential that generates a large $SU(4)$ invariant coupling for the Higgs sector and only small set of particles charged under $SU(2)_X$, which allows the model to be perturbative around the Planck scale. The new gauge interaction drives the top yukawa coupling small at higher energy scales, which also reduces the tuning.
}

\end{center}
\end{titlepage}

\setcounter{page}{2}

\section{Introduction}

Supersymmetry (SUSY) provides one of the most promising solutions to the hierarchy problem of the Standard Model (SM)~\cite{MaianiLecture,Veltman:1980mj,Witten:1981nf,Kaul:1981wp}. However, the lack of finding of SUSY partners casts serious doubts on whether SUSY can still naturally explain the electroweak (EW) scale. Fine-tuning of the EW scale in minimal SUSY models implied by the LHC searches was recently quantified in refs.~\cite{Buckley:2016kvr,Buckley:2016tbs}, which demonstrated that the current limits on stop and gluino masses exclude regions with fine-tuning better than 10$\%$, even if a very low mediation scale of the SUSY breaking of 100 TeV is assumed.\footnote{The fine-tuning may be improved if the higgsino mass is not tied to the Higgs mass squared, see e.g.~refs.~\cite{Dimopoulos:2014aua,Cohen:2015ala,Martin:2015eca,Garcia:2015sfa}. In such a case higgsino could be heavier leading to compressed spectra for which the lower bounds on stops, and gluino are much weaker.}  The fine-tuning quickly 
gets worse for larger mediation scales due to longer RG running of the soft Higgs mass.
 This is  indication of the little hierarchy problem.

A possible remedy to  the little hierarchy problem is offered by Twin Higgs mechanism~\cite{Chacko:2005pe,Chacko:2005vw,Chacko:2005un,Falkowski:2006qq,Chang:2006ra}. In the scenario, the Higgs is a pseudo-Nambu-Goldstone boson of a global $SU(4)$ symmetry emerging from $\mathbb{Z}_2$ symmetry exchanging the SM with its mirror (or twin) copy.
We refer to~\cite{Batra:2008jy,Geller:2014kta,Barbieri:2015lqa,Low:2015nqa,Cheng:2015buv,Csaki:2015gfd,Cheng:2016uqk,Contino:2017moj} for composite Twin Higgs models, and~\cite{Barbieri:2016zxn,Chacko:2016hvu,Craig:2016lyx,Garcia:2015loa,Garcia:2015toa,Craig:2015xla,Farina:2015uea,Freytsis:2016dgf,Farina:2016ndq,Prilepina:2016rlq,Barbieri:2017opf} for cosmological aspects of Twin Higgs scenario. 

Early realisations of  SUSY UV completion of Twin Higgs scenario~\cite{Falkowski:2006qq,Chang:2006ra}, which generate an $SU(4)$ invariant quartic term with an $F$-term potential of a heavy singlet superfield, are not able to significantly reduce fine-tuning as compared to non-Twin SUSY models~\cite{Craig:2013fga,Katz:2016wtw,Badziak:2017syq}. 
It was only very recently that SUSY Twin Higgs models were proposed in which tuning at the level of 10$\%$ is possible
by  introducing either hard $\mathbb{Z}_2$ symmetry breaking in the $F$-term model~\cite{Katz:2016wtw} or a new $U(1)_X$ gauge symmetry whose $D$-term potential provides a large $SU(4)$ invariant quartic term~\cite{Badziak:2017syq}. It should be, however, emphasised that the tuning at the level of 10$\%$ can be obtained in these models only for a low mediation scale or a low Landau pole scale.
In the $F$-term model of ref.~\cite{Katz:2016wtw}
a fine-tuning penalty for a larger mediation scale and hence a longer RG running is severe because the large $SU(4)$ invariant coupling induces growth of the top yukawa coupling at higher energy scales.
In the $D$-term model the RG effect of the gauge coupling $g_X$ of the new interaction is to reduce the top yukawa coupling, and the effect of a higher mediation scale is not as severe as the one for the $F$-term model. However, the RG running of the $U(1)_X$ gauge coupling is fast 
and hence the Landau pole scale of $g_X$ is as low as $10^5-10^6$ GeV for values of $g_X$ that are large enough to guarantee approximate $SU(4)$ symmetry of the Higgs potential.
While such a low mediation scale or a low Landau pole scale is in principle possible,
it strongly limits possible schemes of the mediation of the SUSY breaking and UV completions above the Landau pole scale.

In the present work, we point out that the Landau pole scale and the mediation scale of the $D$-term model can be much higher 
if the $SU(4)$ invariant term is generated by a $D$-term potential of a new non-abelian gauge symmetry. We construct a consistent model with $SU(2)_X$
gauge symmetry with small number of flavors charged under this symmetry.
The new gauge interaction drives the top yukawa coupling small at higher energy scales, which also helps obtain the EW scale more naturally.
As a result, the tuning of the EW scale for 2 TeV stops and gluino can be at
the level
of $5-10\%$ for mediation scales as high as $10^9-10^{13}$ GeV.
One can keep perturbativity up to around the Planck scale with tuning better than $5\%$ (for low mediation scales). 
The model allows for moderate tuning better than few percent with the mediation scale around the Planck scale.
If the gluino mass is a Dirac one, the tuning may be as good as $10\%$, which realizes a natural SUSY with a gravity mediation.

\section{A SUSY $D$-term Twin Higgs with an $SU(2)$ gauge symmetry}

In this section we present a SUSY $D$-term Twin Higgs model~\cite{Badziak:2017syq} where the $D$-term potential of a new $SU(2)_X$ gauge symmetry
generates the $SU(4)$ invariant quartic coupling.
We assume a $\mathbb{Z}_2$ symmetry exchanging the SM with its mirror copy, and denote mirror objects with supersctripts $'$.

The matter content of the model is shown in Table~\ref{tab:matter}.
In addition to the $SU(3)_c\times SU(2)_L\times U(1)_Y$ gauge symmetry and its mirror counterpart, we introduce an $SU(2)_X$ gauge symmetry which is neutral under the $\mathbb{Z}_2$ symmetry.
We embed an up-type Higgs $H_u$ into a bi-fundamental of $SU(2)_L\times SU(2)_X$, ${\cal H}$, and its mirror partner $H_u'$ into that of $SU(2)'_L\times SU(2)_X$, ${\cal H}'$. 
As we will see later, the $D$-term potential of $SU(2)_X$ is responsible for the $SU(4)$ invariant quartic coupling of $H_u$ and $H_u'$.
The $SU(2)_X$ symmetry is broken by the vacuum expectation value (VEV) of a pair of $SU(2)_X$ fundamental $S$ and $\bar{S}$.
Except for $S$ and $\bar{S}$ all matter fields have their mirror partner.

The right-handed top quark is embedded into $\bar{Q}_R$ and allow for a large enough top yukawa coupling through the superpotential term ${\cal H}\bar{Q}_R Q_3$, where $Q_3$ is the third generation quark doublet. $\bar{E}$ is necessary in order to cancel the $U(1)_Y\mathchar`-SU(2)_X^2$ anomaly.
The VEV of $\phi_u$ is responsible for the masses of the up and charm quarks.
$Q_{1,2,3}$, $\bar{u}_{1,2}$, $\bar{e}_{1,2,3}$, $\bar{d}_{1,2,3}$ and $L_{1,2,3}$ are usual MSSM fields.
To cancel the gauge anomaly of 
$SU(3)_c^2 \mathchar`-U(1)_Y$ and $U(1)_Y^3$
originating from the extra up-type right handed quark in $\bar{Q}_R$ and two extra right-handed leptons in $\bar{E}$, we introduce $U$ and $E_{1,2}$.
There are three up-type Higgses in ${\cal H}$ and $\phi_u$, so we need to introduce three down-type Higgsses $\phi_{d1,2,3}$.
Their VEVs are responsible for the masses of down-type quarks and charged leptons.

\begin{table}[htp]
\caption{The matter content of the model.}
\begin{center}
\begin{tabular}{|c|c|c|c|c|c|c|c|}
                          &$SU(2)_X$&$SU(2)_L$&$SU(2)_L'$&$U(1)_Y$&$U(1)_Y'$&$SU(3)_c$    &$SU(3)_c'$   \\ \hline
${\cal H}$          & ${\bf 2}$    & ${\bf 2}$   &                  & $1/2$     &                &                      &                     \\
${\cal H}'$          &  ${\bf 2}$  &                  & ${\bf 2}$   &               & $1/2$       &                     &                     \\
$\bar{Q}_R$      &  ${\bf 2}$  &                  &                  & $-2/3$    &                &${\bf \bar{3}}$&                     \\
$\bar{Q}_R'$      & ${\bf 2}$   &                 &                  &               & $-2/3$     &                      &${\bf \bar{3}}$\\
$S$                    & ${\bf 2}$   &                 &                  &               &                &                      &                      \\
$\bar{S}$            & ${\bf 2}$    &                 &                  &               &                &                      &                      \\
$\bar{E}$            & ${\bf 2}$    &                 &                  & $1$        &                &                      &                      \\
$\bar{E}'$           & ${\bf 2}$    &                 &                  &               & $1$         &                      &                      \\
$U$                     &                  &                 &                  &  $2/3$     &                 & ${\bf 3}$       &                      \\
$U'$                    &                  &                 &                  &               &$2/3$         &                     & ${\bf 3}$        \\
$E_{1,2}$            &                  &                 &                  & $-1$        &                 &                      &                      \\
$E'_{1,2}$            &                  &                 &                  &               &  $-1$        &                      &                      \\
$\phi_u$              &                  & ${\bf 2}$   &                  & $1/2$     &                 &                      &                      \\
$\phi_u'$              &                  &                 & ${\bf 2}$    &               &                 &                      &                      \\
$\phi_{d1,2,3}$     &                  &${\bf 2}$   &                  & $-1/2$    &                 &                      &                      \\
$\phi_{d1,2,3}'$     &                &                 &${\bf 2}$      &               & $-1/2$   &                      &                      \\
$Q_{1,2,3}$          &                 &  ${\bf 2}$   &                 &  $1/6$     &                 &  ${\bf 3}$      &                      \\
$\bar{u}_{1,2}$     &                  &                 &                  & $-2/3$    &                 &${\bf \bar{3}}$&                      \\
$\bar{e}_{1,2,3}$ &                  &                 &                  &  $1$        &                 &                      &                      \\
$\bar{d}_{1,2,3}$ &                  &                 &                  & $1/3$      &                 &${\bf \bar{3}}$&                      \\
$L_{1,2,3}$          &                  &  ${\bf 2}$  &                  & $-1/2$    &                 &                      &                      \\
$Q_{1,2,3}'$          &                 &                 &  ${\bf 2}$   &               & $1/6$       &                      &  ${\bf 3}$       \\
$\bar{u}_{1,2}'$     &                  &                 &                  &               & $-2/3$      &                      & ${\bf \bar{3}}$ \\
$\bar{e}_{1,2,3}'$ &                  &                 &                  &               &  $1$         &                      &                      \\
$\bar{d}_{1,2,3}'$ &                  &                 &                  &               & $1/3$       &                      & ${\bf \bar{3}}$\\
$L_{1,2,3}'$          &                  &                 &  ${\bf 2}$   &               &   $-1/2$    &                      &                      \\ \hline
\end{tabular}
\end{center}
\label{tab:matter}
\end{table}%

\subsection{$SU(2)_X$ symmetry breaking}
We introduce a singlet chiral field $Z$ and the superpotential coupling
\begin{align}
\label{eq:WkappaZSS}
W = \kappa Z (S \bar{S} - M^2).
\end{align}
We assume that the soft masses of $S$ and $\bar{S}$ are the same,
\begin{align}
V_{\rm soft} = m_S^2 (|S|^2 + |\bar{S}|^2).
\end{align}
Otherwise, the magnitude of the VEVs of $S$ and $\bar{S}$ are different from each other, and give large soft masses to the Higgs doublets through the $D$-term potential. The VEVs of $S$ and $\bar{S}$ are given by
\begin{align}
\vev{S} =
\begin{pmatrix}
0 \\ v_S
\end{pmatrix},~
\vev{\bar{S}} =
\begin{pmatrix}
v_S \\ 0
\end{pmatrix},~~v_S = \sqrt{M^2 - m_S^2 / \kappa^2}.
\end{align}
The constraint on the $T$ ($\rho$) parameter requires that $v_S \gtrsim2.9$ TeV in the limit of large $\tan\beta$ and neglecting the effect of mixing between the SM and the mirror Higgses, see Appendix~\ref{sec:EWPM} for a derivation of this constraint and more precise formula.
The masses of the $SU(2)_X$ gauge bosons are given by
\begin{align}
m_X^2 = g_X^2 v_S^2.
\end{align}

After integrating out massive particles with a mass as large as $v_S$, the potential of ${\cal H} $ and ${\cal H}'$ is given by
\begin{align}
\frac{1}{8}g_X^2 \sum_{i=1,2,3} \left( {\cal H}^\dag \sigma^i {\cal H} + {\cal H'}^\dag \sigma^i {\cal H'}\right)^2  \left( 1 - \epsilon^2 \right),\\
\label{eq:epsilon}
\epsilon^2 = \frac{m_X^2}{2m_S^2 + m_X^2}.
\end{align}
In the SUSY limit, $m_S^2 =0$, the $D$-term potential vanishes.
In terms of the model parameters $M,m_S, \kappa, g_X$, $\epsilon^2$ is given by
\begin{align}
\epsilon^2 = \frac{g_X^2 (m_S^2 - \kappa^2 M^2)}{g_X^2 (m_S^2 - \kappa^2 M^2) - 2 \kappa^2 m_S^2 }
\end{align}
In the limit where $\kappa \ll g_X$, $\epsilon^2=1$ and hence the $D$-term potential decouples.
In order to obtain a large $D$-term potential, it is preferable that $\kappa$ is as large $g_X$.

\begin{figure}[tb]
\centering
\includegraphics[clip,width=.48\textwidth]{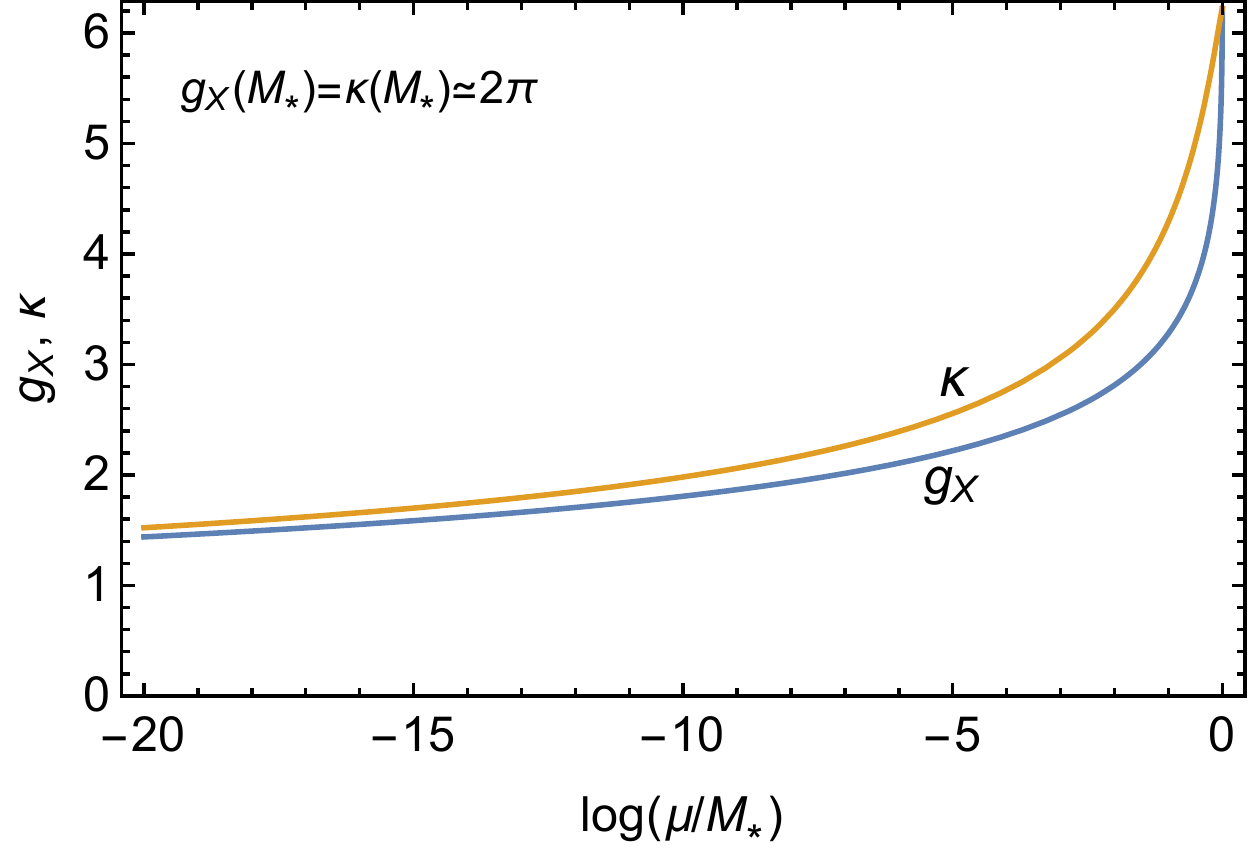}
\caption{Running of $g_X$ and $\kappa$ from a high energy scale $M_*$ down to lower energy scales.
}
\label{fig:kappa_run}
\end{figure}

To estimate the maximal possible value of $\kappa$, we solve the renormalization group equation of $g_X$ and $\kappa$,
\begin{align}
\label{eq:beta_gx}
\frac{{\rm d}}{{\rm dln}\mu} g_X =& \frac{g_X^3}{16\pi^2} \frac{ 1 + \frac{21}{16\pi^2} g_X^2 - \frac{1}{8\pi^2} \kappa^2}{1-\frac{g_X^2}{4\pi^2}},\\
\frac{{\rm d}}{{\rm dln}\mu} \kappa =& \frac{\kappa}{16\pi^2} (4 \kappa^2 - 3 g_X^2),
\end{align}
from a high energy scale $M_*$ towards low energy scales, with a boundary condition at $M_*$ of $g_X = \kappa \simeq 2\pi$.
$M_*$ can be identified with the Landau pole scale. 
The running of $g_X$ and $\kappa$ is shown in Fig.~\ref{fig:kappa_run}, which shows that $\kappa \simeq g_X$ much below $M_*$.
We obtain the same conclusion as long as $\kappa(M_*) \gsim 1$.
For $\kappa\simeq g_X$, $\epsilon^2$ is
\begin{align}
\epsilon^2 \simeq \frac{g_X^2 M^2 - m_S^2}{g_X^2 M^2 + m_S^2}
\end{align}
We may obtain a sufficiently small $\epsilon^2$, say $\epsilon^2 \lsim 0.2$, for $m_S^2 \gsim 0.6 g_X^2 M^2$.

Notice also that for $\epsilon^2<1$ there is a threshold correction to the soft Higgs mass which is proportional to
a new gauge bosons mass squared:
\begin{equation}
\label{deltamHu_X}
 \left(\delta m_{H_u}^2\right)_{X}= 3 \frac{g_X^2}{64 \pi^2} m_X^2 \ln\left(\epsilon^{-2}\right) \,,
\end{equation}
which may be a source of tuning of the EW scale. The same threshold correction is present also for the right-handed stop soft mass squared $m_{U_3}^2$. 

\subsection{$SU(4)$ invariant quartic coupling and $\mu$ terms}

We give masses to ${\cal H} = (H_1,H_2)^T$ and $\phi_u$ by pairing them with $\phi_{d,1,2,3}$ through the superpotential terms,
\begin{align}
\label{eq:Whiggs1}
W = \lambda_1 \phi_{d,1}{\cal H} S + \lambda_2 \phi_{d,2} {\cal H} \bar{S} + m \phi_u \phi_{d,3} 
\end{align}
The pairs $(H_1,\phi_{d,1})$, $(H_2,\phi_{d,2})$ and $(\phi_u,\phi_{d,3})$ obtain masses of $\lambda_1 v_S$, $\lambda_2 v_S$ and $m$, respectively.
We assume that $\lambda_2 v_S, m \gsim 1$ TeV and neglect $(H_2,\phi_{d,2})$ and $(\phi_u,\phi_{d,3})$ for the dynamics of the electroweak symmetry breaking.
We identify $H_1$ and $\phi_{d,1}$ with $H_u$ and $H_d$ in the Higgs sector of the standard SUSY model.
The $\mu$ parameter is given by $\mu=\lambda_1 v_S$. 
The $SU(4)$ invariant quartic coupling of $(H_u,H_u')$ is given by
\begin{align}
V =\frac{g_X^2}{8} (1-\epsilon^2) (|H_u|^2 + |H_u'|^2)^2.
\end{align}

As we will see, the VEV of $\phi_u$ is responsible for the masses of the up and charm quarks, and the neutrinos.
To give a VEV to $\phi_u$, we introduce a coupling
\begin{align}
\label{eq:Whiggs2}
W = \delta m \phi_u \phi_{d,1}.
\end{align}
Through the $F$ term potential of $\phi_{d,1}$, $\phi_u$ obtains a tadpole term after $H_u$ obtains its VEV, which induces a non-zero VEV of $\phi_u$.

Through the coupling $\lambda_2 (> \lambda_1)$, $m_{H_u}^2$ receive a quantum correction from $m_{S}^2$,
\begin{align}
\label{eq:mHu_direct}
\Delta m_{H_u}^2  \simeq - \frac{\lambda_2^2}{8\pi^2}m_S^2 L = - (600~{\rm GeV})^2 \left( \frac{\lambda_2}{0.3} \right)^2 \frac{m_S^2}{(6~{\rm TeV})^2} \frac{L}{{\rm ln}10^4}
\end{align}
where $L$ denotes a log-enhancement through an RGE.
As long as $\lambda_2\lsim 0.4$, this contribution is always smaller than that from stops and/or the threshold correction from $X$, and hence we
neglect it. Note, however, that even larger values of $\lambda_2$ may be possible without introducing tuning if the mediation scale of SUSY breaking
is relatively low and/or $m_S^2$ runs to smaller values at higher energies.

Note that the $Z_2$ symmetry $S \leftrightarrow \bar{S}$ is explicitly broken by the above superpotential couplings.
Even if we assume the $Z_2$ symmetry of the soft masses of $S$ and $\bar{S}$, we expect a quantum correction to a mass difference of them,
\begin{align}
\Delta m_S^2 \equiv m_S^2 - m_{\bar{S}}^2 \simeq \frac{\lambda_2^2}{8\pi^2}m_S^2 L.
\end{align}
This leads a asymmetric VEV of $S$ and $\bar{S}$, which give $m_{H_u}^2$ through the $D$-term potential,
\begin{align}
m_{H_u^2} \simeq -\frac{\epsilon^2}{2} \Delta m_S^2,
\end{align}
which is always smaller than the direct one-loop quantum correction in Eq.~(\ref{eq:mHu_direct}).

It is also possible to maintain the $Z_2$ symmetry.
Instead of the coupling in Eqs.~(\ref{eq:Whiggs1}) and (\ref{eq:Whiggs2}), we introduce
\begin{align}
\label{eq:mu_heavy}
W ={\cal H} ( \lambda_1 \phi_{d,1} + \lambda_3 \phi_{d,3} )( S + \bar{S}) + \lambda_2 {\cal H}\phi_{d,2} ( S - \bar{S})  + \phi_u ( m_1\phi_{d,1} + m_3 \phi_{d,3})
\end{align}
Here we have assumed that $\phi_{d,2}$ is odd under the $Z_2$ symmetry.
After $S$ and $\bar{S}$ obtain their VEVs, the mass terms become
\begin{align}
W = v_S (\lambda_1 \phi_{d,1} + \lambda_3 \phi_{d,3} ) (H_1 - H_2) + \lambda_2 v_S \phi_{d,2} (H_1 + H_2) + \phi_u ( m_1\phi_{d,1} + m_3 \phi_{d,3}).
\end{align}
We assume that $\lambda_2 v_S, m_i \gsim 1$ TeV.
Then $(H_1 + H_2) /\sqrt{2}$ and $\phi_u$ obtain a large mass paired with  $\phi_{d,2}$ and a linear combination of $\phi_{d,1}$ and $\phi_{d,3}$, respectively, and 
are irrelevant for the dynamics of the electroweak symmetry breaking.
$H_u \equiv (H_1 - H_2)/\sqrt{2}$ obtains a mass of $O(\lambda_{2,3} v_S)$ paired with another linear combination of $\phi_{d,1}$ and $\phi_{d,3}$ which we call $H_d$.

\subsection{Masses of matter particles}

We first consider a case where the $Z_2$ symmetry $S\leftrightarrow \bar{S}$ is explicitly broken.
A large enough top yukawa coupling is obtained by the superpotential
\begin{align}
W = y_t {\cal H} \bar{Q}_R Q_3 \rightarrow  y_t  \left(H_2\bar{Q}_{R,1} - H_1 \bar{Q}_{R,2}  \right) Q_3,
\end{align}
where $\bar{Q}_R = ( \bar{Q}_{R,1} , \bar{Q}_{R,2})^T$.
We give a large mass to $\bar{Q}_{R,1}$ by introducing a coupling
\begin{align}
W = y \bar{Q}_R U S,
\end{align}
and identify $\bar{Q}_{R,2}$ with a right-handed top quark $\bar{u}_3$.

The yukawa couplings of the up and charm quarks originates from the couplings with $\phi_u$,
\begin{align}
W = y_{u,ij}\phi_u Q_i \bar{u_j}.
\end{align}
The left-handed neutrino masses are obtained in a similar manner once right-handed neutrinos are introduced.
The yukawa couplings of the down-type quarks and the charged leptons is given by couplings with $\phi_{d,i}$,
\begin{align}
W = y_{d,ijk}\phi_{d,i} Q_j \bar{d}_k + y_{e,ijk}\phi_{d,i} L_j \bar{e}_k.
\end{align}

The extra $SU(2)_X$ charged particle $\bar{E}$ obtains its mass paired with $E_{1,2}$ through the $SU(2)_X$ symmetry breaking,
\begin{align}
W  = \bar{E} (y_{E,1} E_1 + y_{E,2} E_2 ) S +  \bar{E} (\bar{y}_{E,1} E_1 + \bar{y}_{E,2} E_2 ) \bar{S}.
\end{align}

Next we consider a case where the $Z_2$ symmetry is maintained. The top yukawa coupling is obtained by the superpotential
\begin{align}
W = y_t {\cal H} \bar{Q}_R Q_3 \rightarrow  y_t H_u \frac{1}{\sqrt{2}}  \left(\bar{Q}_{R,1} +  \bar{Q}_{R,2}  \right) Q_3.
\end{align}
One linear combination of $\bar{Q}_{R,1}$ and $\bar{Q}_{R,2}$ obtains a Dirac mass term paired with $U$,
\begin{align}
\label{eq:top yukawa}
W = y \bar{Q}_R U (S + \bar{S}) \rightarrow y v_S \frac{1}{\sqrt{2}} ( \bar{Q}_{R,1} - \bar{Q}_{R,2} )U.
\end{align}
We identify the massless combination $(\bar{Q}_{R,1} +\bar{Q}_{R,2} )/\sqrt{2} \equiv \bar{u}_3$ as a right-handed top quark.
The extra $SU(2)_X$ charged particle $\bar{E}$ obtains its mass paired with $E_{1,2}$ through the coupling,
\begin{align}
W  = \bar{E} (y_{E,1} E_1 + y_{E,2} E_2 ) S +  \bar{E} (y_{E,1} E_1 - y_{E,2} E_2 ) \bar{S}.
\end{align}
Here we assume that $E_2$ is odd under the $Z_2$ symmetry $S \leftrightarrow \bar{S}$, so that all particles in $\bar{E}$ and $E_{1,2}$ obtains their masses.

So far we have assumed that a linear combination of $\bar{Q}_{R,1}$ and $\bar{Q}_{R,2}$ obtains a large mass paired with $U$.
It is also possible to identify the linear combination with the right-handed charm quark. In such a model $U$ and $\bar{u}_2$ are not necessary.
The mass of the right-handed scharm is predicted to be as large as that of the right-handed stop.
This choice is beneficial for a high mediation scale, as it makes the $SU(3)_c$ and $U(1)_Y$ coupling constants relatively smaller, reducing the fine-tuning from the gluino and the bino.

\section{Fine-tuning of the electroweak scale}
\label{sec:FT}

Let us now discuss fine-tuning of the EW scale in the model. We quantify the degree of fine-tuning by introducing the measure~\cite{Craig:2013fga},
\begin{equation}
\label{eq:Delta_v}
\Delta_v \equiv   \Delta_f \times \Delta_{v/f},
\end{equation}
where
\begin{align}
\label{eq:Delta_vf}
&\Delta_{v/f} = \frac{1}{2} \left( \frac{f^2}{v^2} -2\right), \\
\label{eq:Delta_f}
&\Delta_f =  {\rm max}_i \left( |\frac{\partial{\rm ln} f^2}{\partial{\rm ln} x_i(\Lambda)}|, 1 \right) .
\end{align}
Here $f \equiv \sqrt{v^2 + v^{'2}}$ is the decay constant of the spontaneous $SU(4)$ breaking. $\Delta_{v/f}$ measures the fine-tuning to obtain $v < f$ via explicit soft $\mathbb{Z}_2$ symmetry breaking. $\Delta_f$ measures the fine-tuning to obtain the scale $f$ from the soft SUSY breaking which is analogous to the fine-tuning to obtain the electroweak scale from the soft SUSY breaking in the MSSM.
$x_i(\Lambda)$ are the parameters of the theory evaluated at the mediation scale of the SUSY breaking $\Lambda$.
We include the important seven parameters, $m_{H_u}^2$, $m_{Q_3}^2$, $m_{\bar{u}_3}^2$, $M_1^2$, $M_2^2$, $M_3^2$ and $\mu^2$.
To evaluate $\Delta_f$ we solve the renormalization group equations (RGEs) of parameters between $\Msusy$ and $\Lambda$.
We assume that the right-handed charm quark is also embedded in $\bar{Q}_R$.
Between $\Msusy$ and $m_X$ we solve MSSM RGEs at the one-loop level
appropriately modifying the beta function of $m_{Q_3}^2$. 
At a scale $m_X$ we perform matching by including the threshold correction \eqref{deltamHu_X} to $m_{H_u}^2$ and $m_{U_3}^2$. 
Above $m_X$ we solve the RGEs (that include the effects of non-MSSM states) at least at the one-loop level. The RGEs of the gauge couplings are solved
at the two-loop level,
but set, for simplicity, $\kappa=0$.\footnote{Non-zero $\kappa$ slightly slows down the running of $g_X$ but the impact on $\Delta_v$ and the scale of  the Landau pole is negligible. }
The yukawa couplings other than the top yukawa are neglected.

As clearly seen from eqs.~\eqref{eq:Delta_v}-\eqref{eq:Delta_f},  for a given value of $f$ there is a lower bound on $\Delta_v$ of $\Delta_{v/f}$. $f/v$ is constrained by the Higgs coupling measurements \cite{Higgscomb} to be at least $2.3$ \cite{Buttazzo:2015bka}. The latter value has been obtained neglecting invisible decays of the Higgs to mirror particles, which are generically non-negligible, so in our numerical analysis we use less extremal value of  $f=3 v$. Nevertheless, the tuning is quite independent of this choice (unless $f$ is so large that $\Delta_{v/f}$ determines $\Delta_v$).

\begin{figure}[p]
\centering
\includegraphics[clip,width=.48\textwidth]{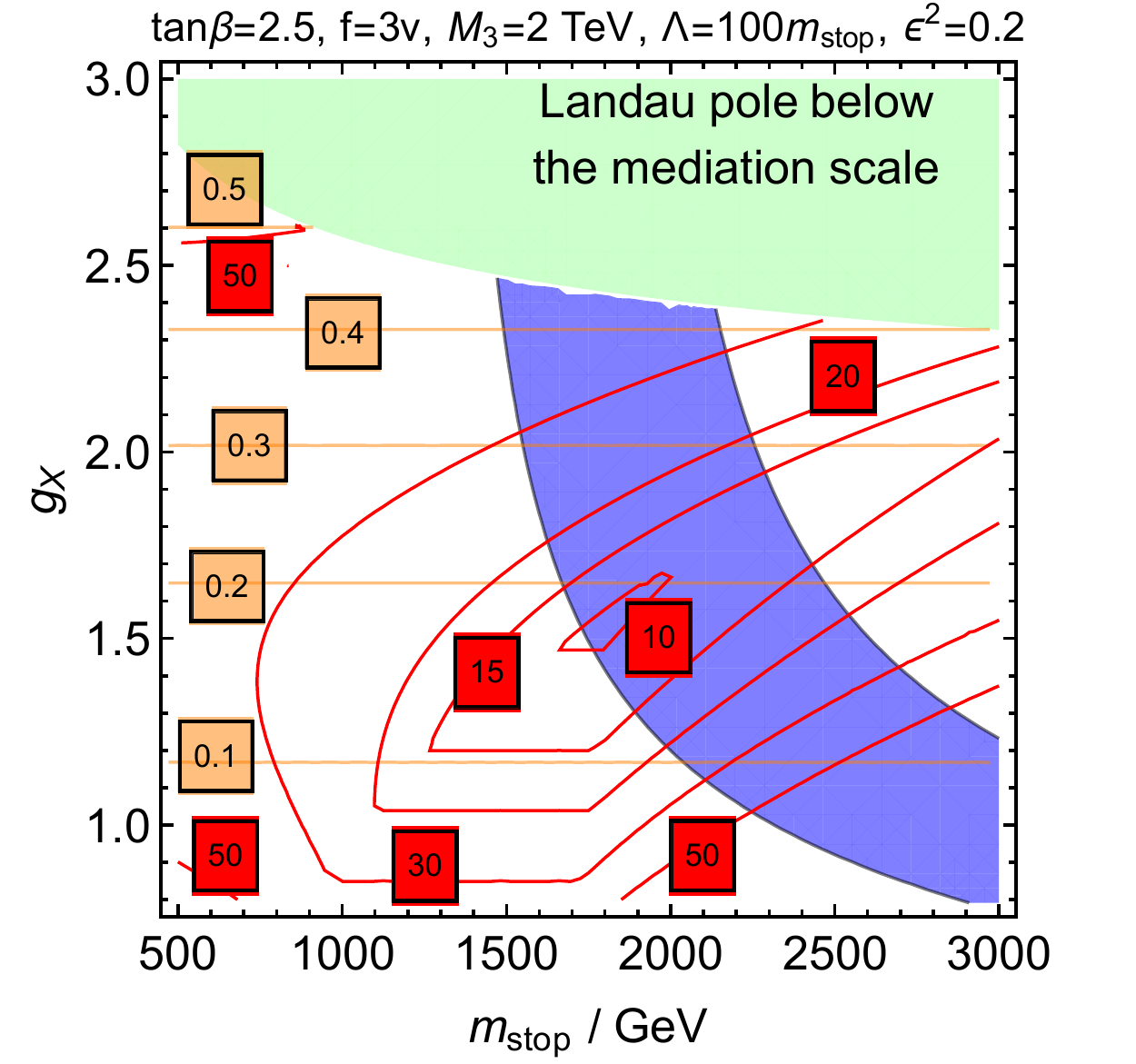}
\includegraphics[clip,width=.48\textwidth]{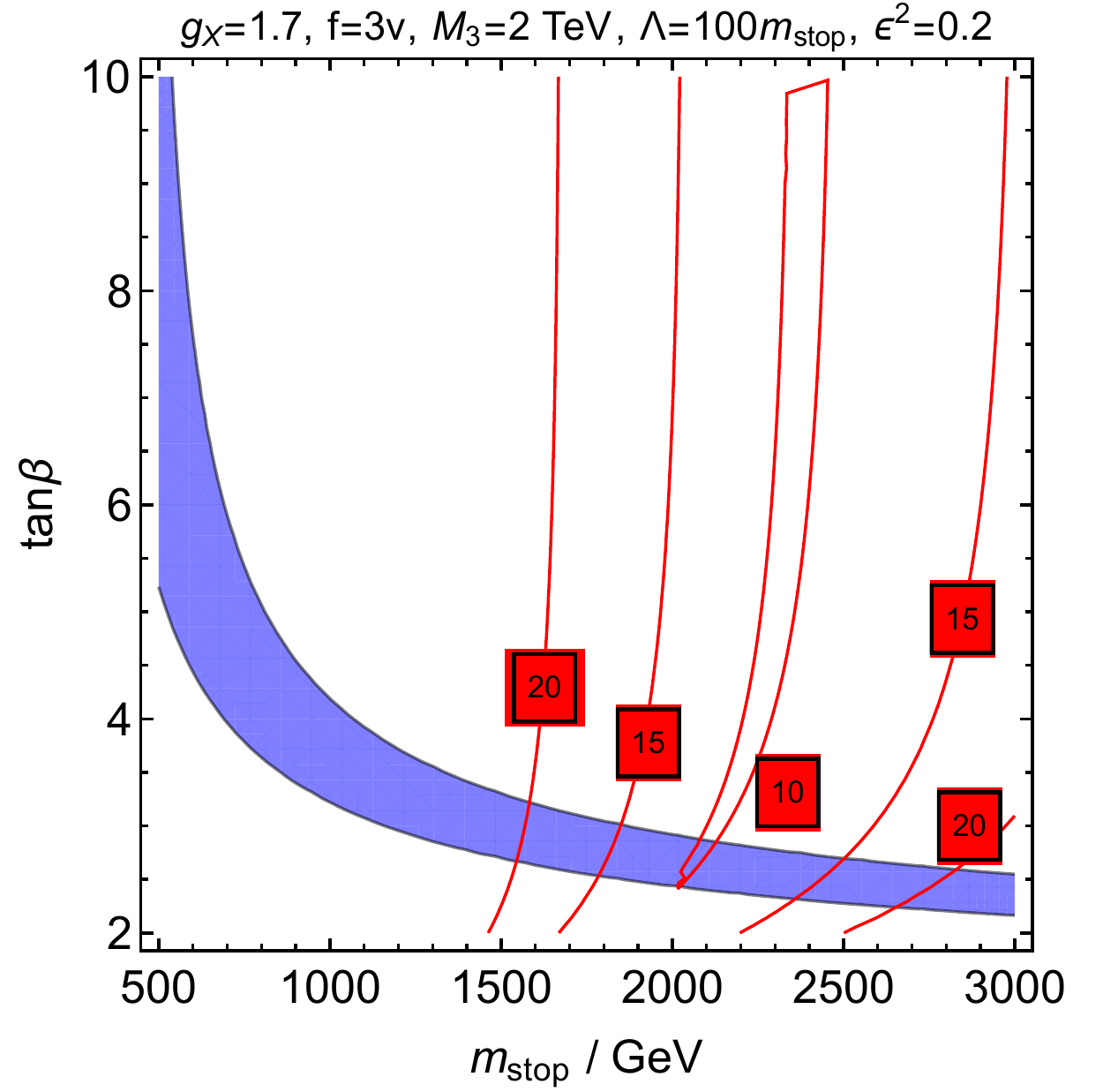}
\includegraphics[clip,width=.48\textwidth]{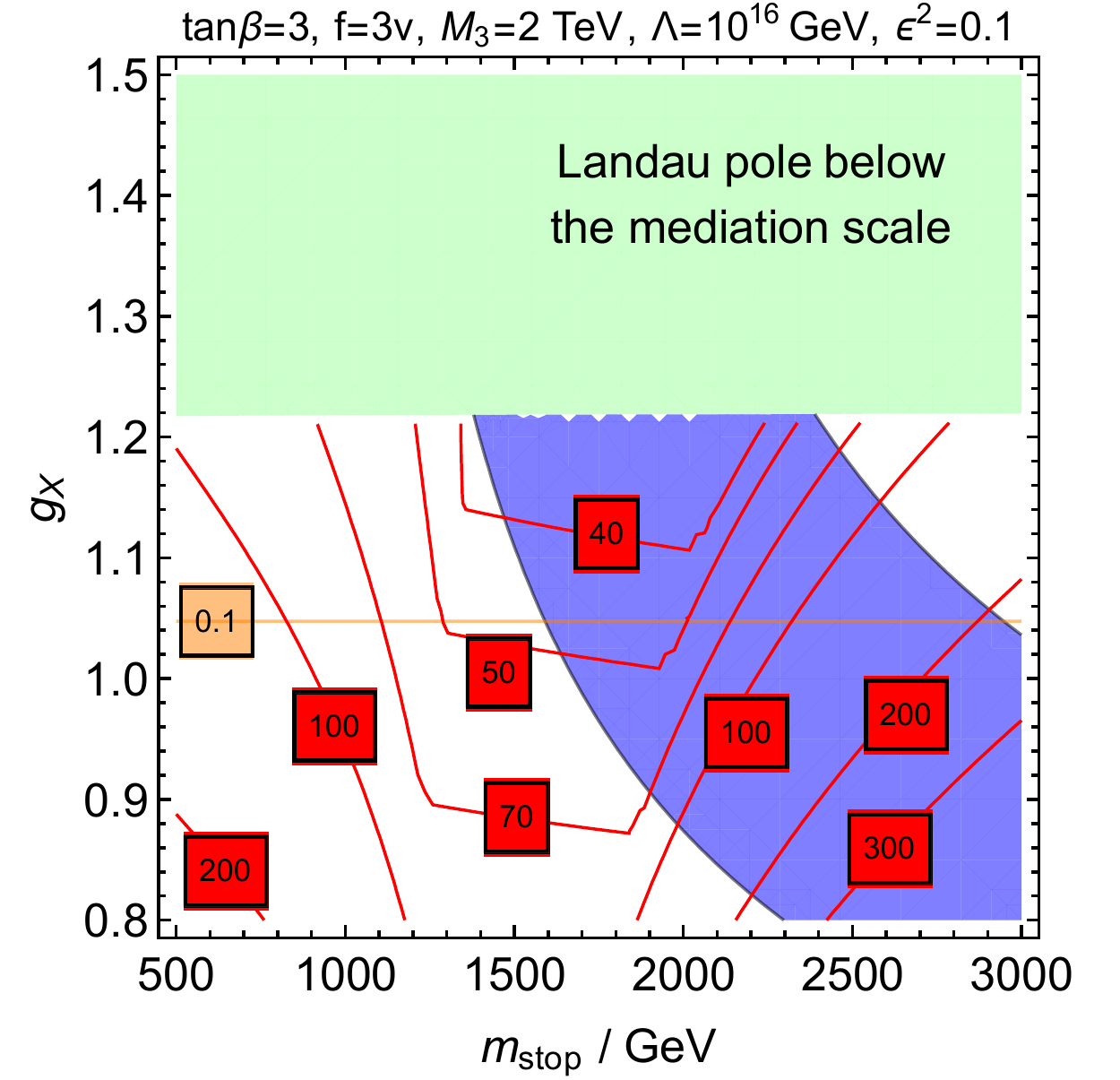}
\includegraphics[clip,width=.48\textwidth]{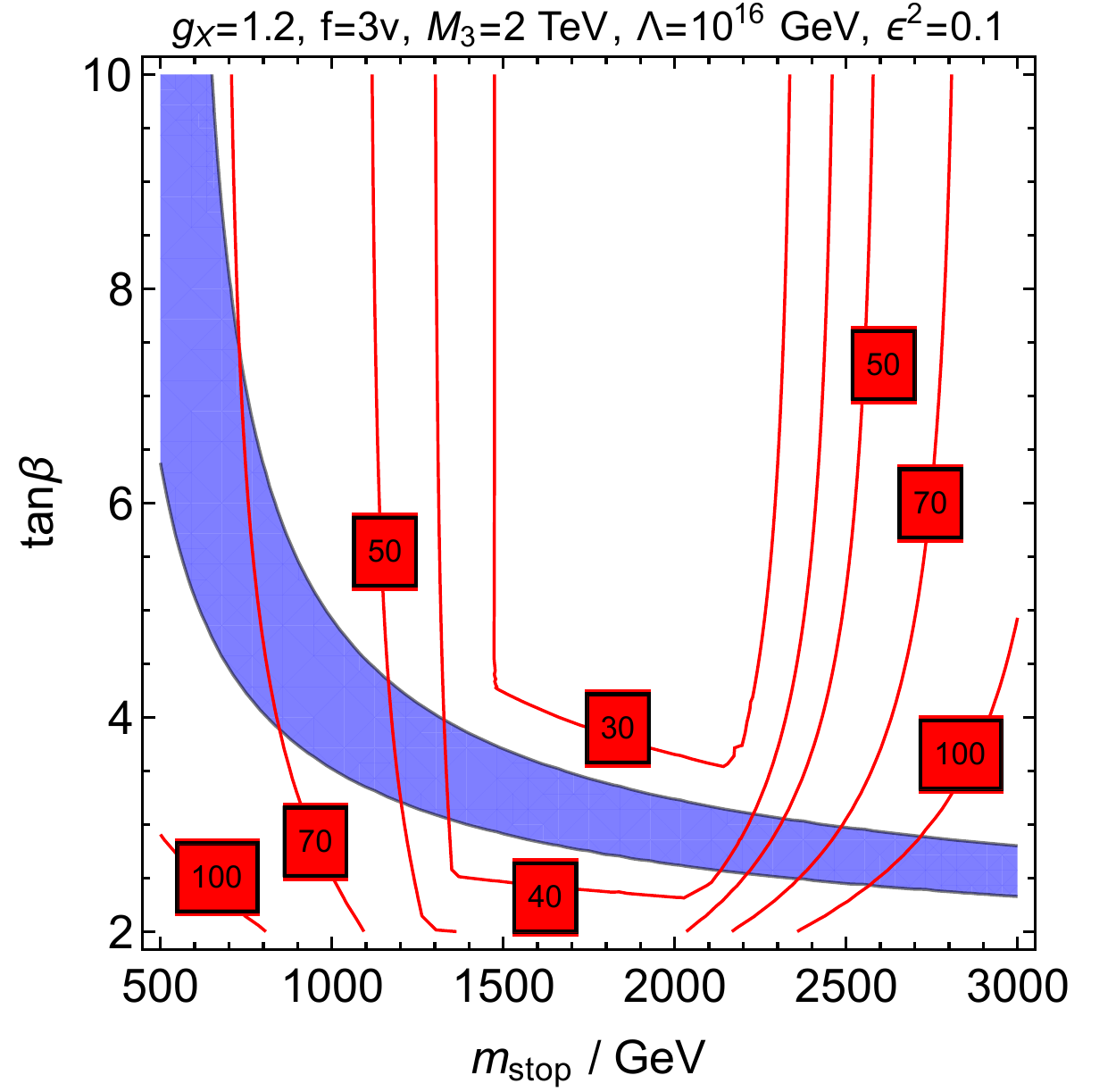}
\caption{Fine-tuning (red contours) in the model for $f=3v$, $\mu=M_1=M_2=500$ GeV, $m_A=1$ TeV and the soft gluino mass term $M_3=2$ TeV  assuming
the mediation
scale $\Lambda=100 \Msusy$ (upper panels) and $\Lambda=10^{16}$ GeV (lower panels). In the left panels, the orange contours depict the value of the $SU(4)$ preserving quartic coupling
and in the green regions the Landau pole of the $SU(2)_X$ gauge coupling constant is below $\Lambda$. In the upper (lower) left panels,
$\tan\beta=2.5$ (3) so that the correct Higgs mass $m_h=125\pm3$~GeV
(the blue region) is obtained for stop masses close to 2 TeV for the most interesting range of $g_X$.  In the right panels, the fine-tuning is shown
in the plane
$\Msusy$-$\tan\beta$ for some fixed values of $g_X$. $m_X$ is chosen such that the  constraint from EW precision measurements is saturated - see Appendix for details.}
\label{fig:ft}
\end{figure}

In fig.~\ref{fig:ft} we present contours of $\Delta_v$ assuming low and high mediation scales of SUSY breaking $\Lambda$.%
\footnote{In the figure we shade the parameter region where the Landau pole scale of the gauge coupling $g_X$ is above $\Lambda$. It is also possible that the SUSY breaking is mediated above the Landau pole scale, but we cannot calculate the fine-tuning measure unless we specify the description of the model above the Landau pole scale.}
Here and hereafter, the stop mass $m_{\rm stop}$ and the gluino mass $M_3$ refer to the values at the TeV scale.
For $\Lambda=100 \Msusy$ tuning at the level of 10$\%$ can be obtained for the stop masses as large as 3 TeV, as seen from the upper left panel. An
important constraint on the parameter space is provided by the Higgs mass measurement \cite{Higgsmass_exp}. In order to assess the impact of this
constraint we compute the Higgs mass following closely the procedure described in ref.~\cite{Badziak:2017syq}. The blue bands show the parameter region with $m_h = 125 \pm 3$ GeV, where the error is a theoretical one. It can be seen from the upper right
panel of fig.~\ref{fig:ft} that this constraint prefers rather light stop unless $\tan\beta$ is small enough.  Since we are most interested in stop
masses that easily avoid current or even potential future LHC constraint we set for the low scale mediation case $\tan\beta=2.5$ which  implies the
stop masses in the range between about 1.5 and 3 TeV. This range narrows to between 1.7 and 2 TeV if one demands tuning better than $10\%$.
Interestingly, tuning is minimised for intermediate values of the stop masses which is a consequence of some cancellation between the threshold 
correction from $X$ and corrections from stops and gluino to $m_{H_u}^2$. In this region the value of $|m_{H_u}^2|$ at the mediation scale is somewhat
suppressed. 
For lighter stops (which can be compatible with the Higgs mass constraint for larger $\tan\beta$) the tuning is dominated by the threshold correction 
which implies tuning at the level of few percent. 
It should be noted that fine-tuning of the EW scale is minimized at some intermediate value of $g_X$ of about $1.5-2$ even though perturbativity
constraint allows for $g_X$ as large as about 2.5. This is because for appropriately large $g_X$ the tuning is dominated by the threshold correction
to $m_{H_u}^2$ from the new gauge bosons. Since the latter must be rather heavy for large $g_X$ due to EW precision constraints, the threshold
correction dominates for $g_X\gtrsim2$ and the tuning gets worse with increasing $g_X$ in spite of larger $SU(4)$ invariant coupling. In fact, for very
large value of $g_X$ there is essentially no tuning of the EW scale from stops and gluino but the overall tuning is at the level of few percent. In
the region of large $g_X$, where the threshold correction dominates the fine-tuning, larger values of $\epsilon$ lead to smaller tuning. On the other
hand, for smaller $g_X$, when the threshold correction is subdominant, it is preferred to have smaller $\epsilon$ to suppress corrections 
from stops and gluino by larger $SU(4)$ invariant coupling.

It is interesting to compare the fine-tuning of the present model to that in the model where an $SU(4)$ invariant coupling originates from a
non-decoupling $D$-term of $U(1)_X$ gauge symmetry proposed in ref.~\cite{Badziak:2017syq}. For the stop mass below about 1 TeV, the $U(1)_X$ is less
tuned with tuning even better than 20$\%$. This is because the threshold correction from the $X$ gauge bosons  in the $U(1)_X$ case is three times
smaller than in the case of $SU(2)_X$. As the stop mass increases the tuning in the $U(1)_X$ model gets worse and already for 2 TeV stops the tuning
in the $SU(2)_X$ model becomes better than in the  $U(1)_X$ model due to larger $SU(4)$ invariant coupling which suppresses the correction from stops.

\begin{figure}[t]
\centering
\includegraphics[clip,width=.48\textwidth]{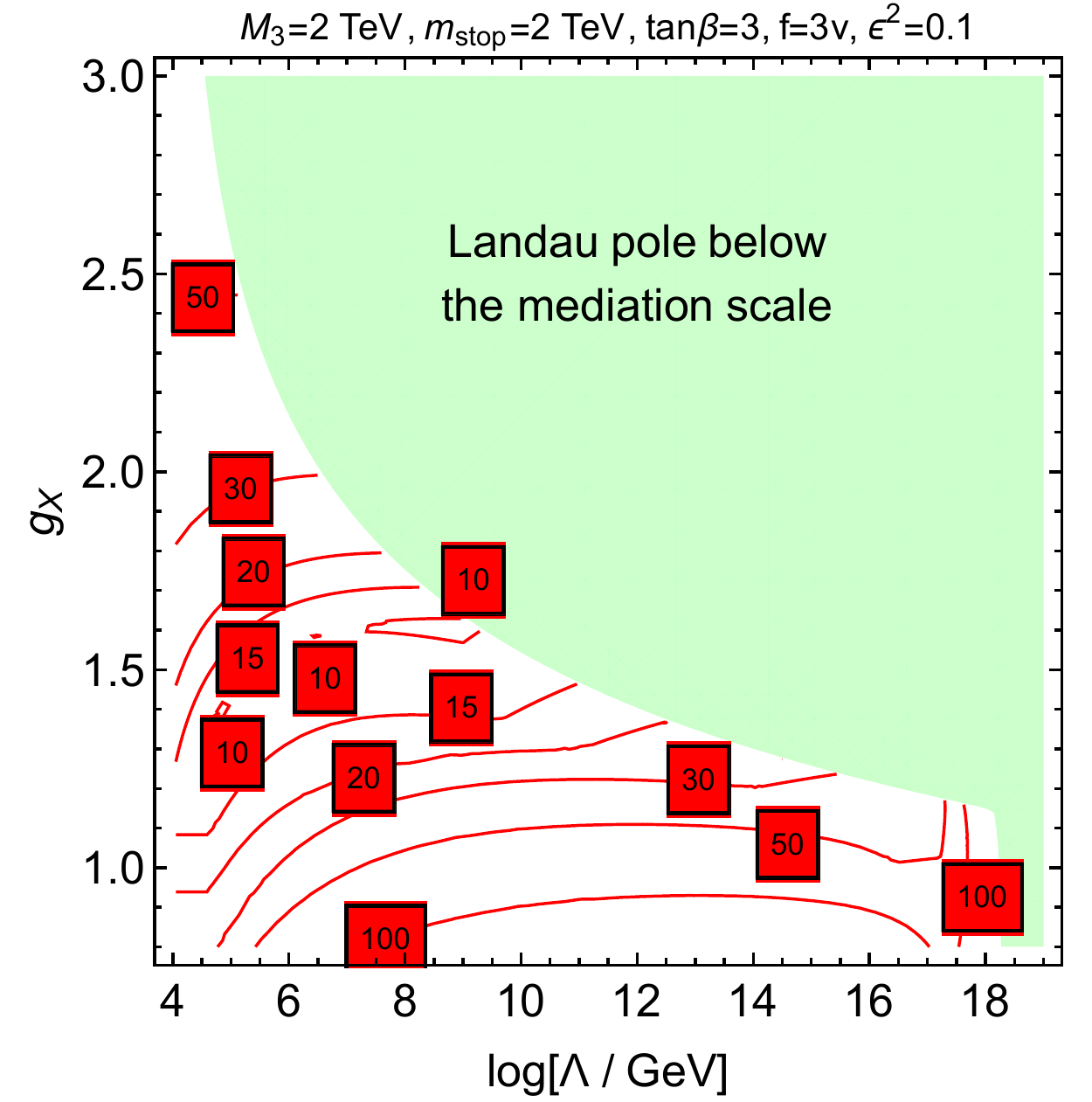}
\includegraphics[clip,width=.48\textwidth]{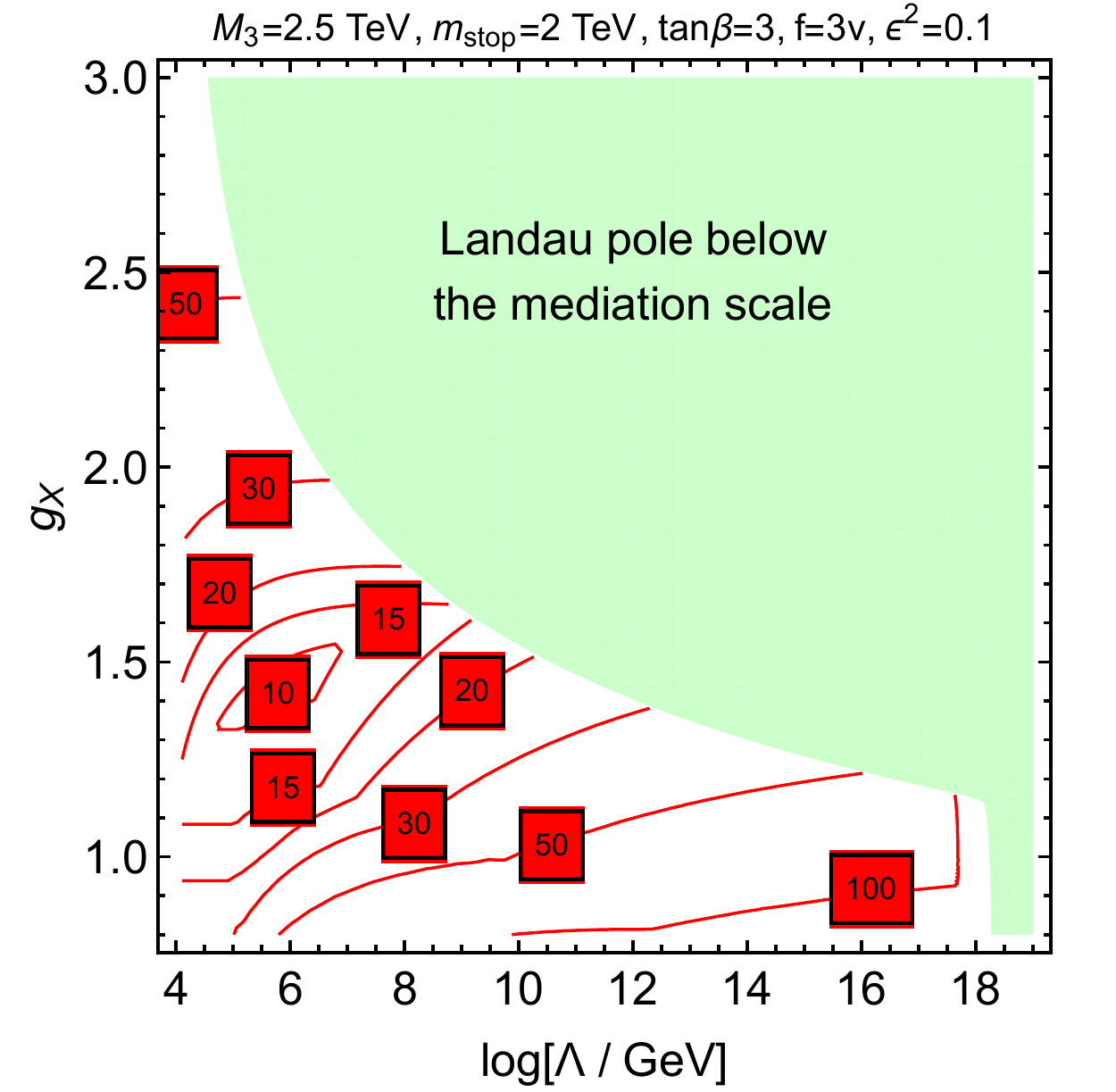}
\caption{
The same as in fig.~\ref{fig:ft} but in the plane $\Lambda$-$g_X$ for $\Msusy=2$~TeV, $\tan\beta=3$ and $M_3=2$~TeV (left panel) or 2.5~TeV (right
panel).  For the chosen values of $\Msusy$ and $\tan\beta$, the Higgs mass is in agreement with the measured value within theoretical uncertainties
in the most of parameter space. For $g_X\gtrsim1.5$ the Higgs mass is slightly too big which can be compensated by reducing $\tan\beta$ by about
10~$\%$ which would have negligible impact on fine-tuning.
}
\label{fig:gxLambda}
\end{figure}

The biggest advantage of the $SU(2)_X$ model is that RGE running of $g_X$ is relatively slow so the Landau pole scale, for given $g_X$, is much higher
than in the  $U(1)_X$ model. 
For example in the case of $\Lambda=10^{16}$ GeV presented in the lower panels of fig.~\ref{fig:ft}, values of $g_X$ up to about 1.2 are possible
without the Landau pole below $\Lambda$.  In the previously proposed SUSY Twin Higgs models it is was not possible to keep perturbativity up to such
high scale. We see from fig.~\ref{fig:ft} that for $\Lambda=10^{16}$ GeV the fine-tuning  better than few~$\%$ can be obtained for the stop masses as
large as 2 TeV. This is obviously worse than in the low-scale mediation case discussed before but for high-scale mediation there are more possible
mechanisms of the mediation of the SUSY breaking. The fine-tuning is also much better than in the MSSM  with high-scale mediation. This is due to
suppression of the corrections from stops and gluino (which dominates tuning for high mediation scales) by 
the $SU(4)$ invariant coupling but also because a large value of $g_X$ efficiently drives the top yukawa coupling to smaller values at higher scales. 
Dependence of fine-tuning on the mediation scale for 2 TeV stops is presented in fig.~\ref{fig:gxLambda}. 
We see that moderate tuning of few percent can be obtained for high mediation scales. For high mediation scales the tuning is dominated by the
correction from the gluino so the tuning crucially depends on the gluino mass limits. It was recently emphasised in ref.~\cite{Buckley:2016tbs}  that
one should convert running soft masses to pole masses when assessing the impact of experimental constraints on naturalness of SUSY models. It was
shown that the loop corrections~\cite{Pierce:1996zz} from 2 TeV squarks increase the gluino pole mass by $10\%$ as compared to the soft mass.  For
heavier 1st/2nd generation of squarks, as experimentally preferred, the correction may be much larger e.g.~20\% for 10 TeV squarks.
In the left panel of fig.~\ref{fig:gxLambda} we fix the soft gluino mass to 2 TeV which easily satisfies the LHC constraints even for moderate loop
corrections from squarks \cite{Sirunyan:2017cwe,ATLAS:2017vjw}.
In such a case, $5\%$ tuning is possible with the mediation scale, being below the Landau pole scale, as high as  $\mathcal{O}(10^{12})$~GeV. For
$M_3=2.5$ TeV, presented in the right panel, for which the gluino is definitely outside of the LHC reach~\cite{Kim:2016rsd}, mediation scale of order
$\mathcal{O}(10^{10})$~GeV can still allow for better than $5\%$ tuning. Notice also a sharp increase in tuning when the mediation scale approaches the
Planck scale. This originates from the fact that $U(1)_Y$ gauge coupling constant runs rather fast due to many new states carrying hypercharge and
eventually enters non-perturbative regime around the Planck scale. In consequence, bino strongly dominates fine-tuning when the mediation scale is
close to the Landau pole for $U(1)_Y$.  

\begin{figure}[t]
\centering
\includegraphics[clip,width=.48\textwidth]{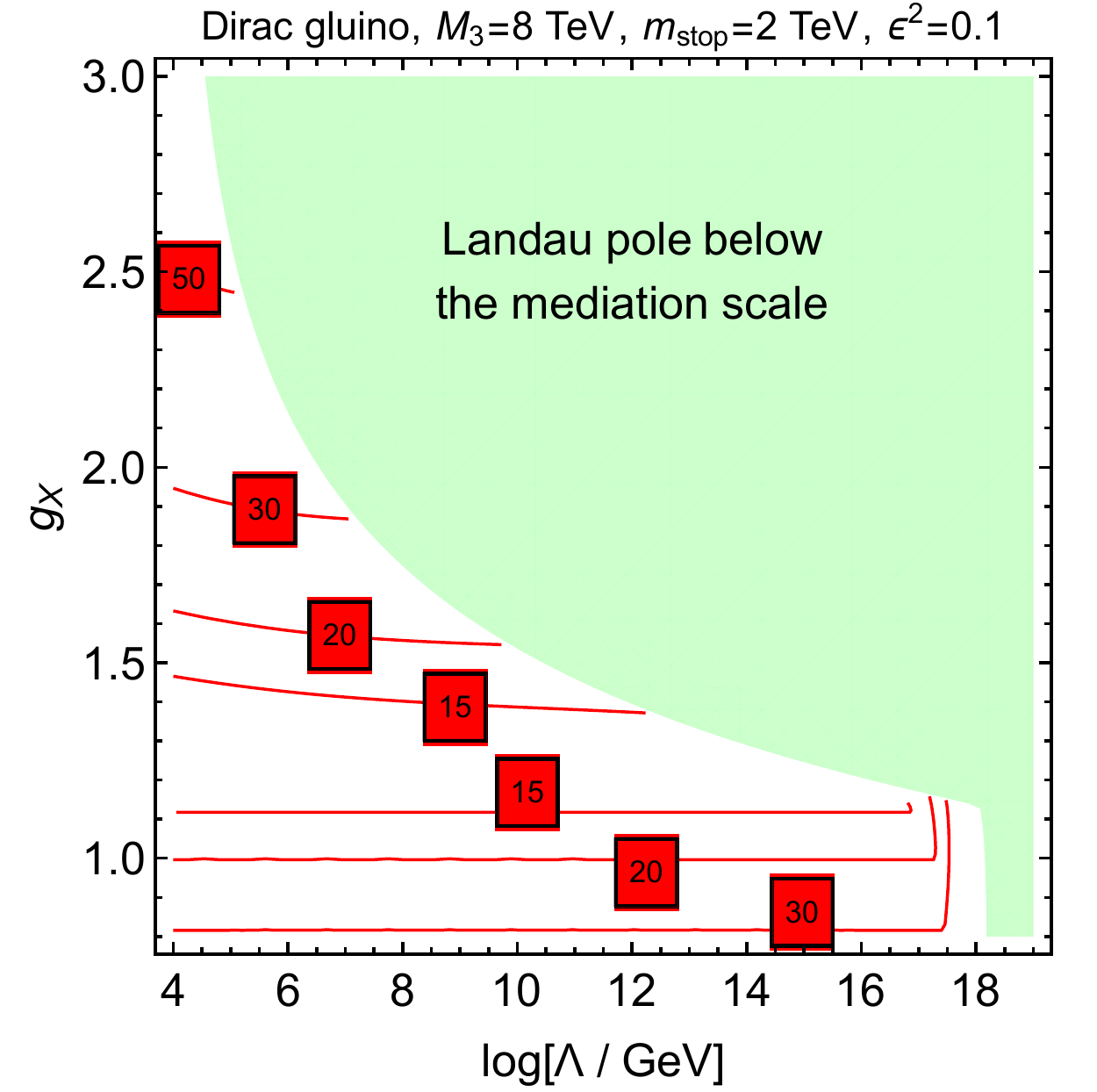}
\caption{
The same as in the left panel of fig.~\ref{fig:gxLambda} but with a Dirac gluino with a soft mass $M_3=8$ TeV and $M_1=M_2=200$ GeV. 
}
\label{fig:dirac}
\end{figure}

The fine-tuning for high mediation scales is even better if the gluino obtains its mass paired with an adjoint chiral superfield by a supersoft
operator, 
due to the absence of the log-enhanced correction to $m_{H_u}^2$~\cite{Fox:2002bu}.
The soft stop mass and the higgs mass are dominantly generated by the threshold correction around the gluino mass,
\begin{align}
m_{\rm stop}^2 \simeq& \frac{1}{16}M_3^2, \\
m_{\rm H_u}^2 \simeq& \frac{3y_t(M_3)^2 }{4\pi^2} m_{\rm stop} {\rm ln}\frac{M_3}{m_{\rm stop}} \,.
\end{align}
In non-Twin models the fine-tuning may be at a few \% level even if the stop mass is as large as 2 TeV, which is further improved by the Twin-Higgs
mechanism.
The contour of $\Delta_v$ assuming the Dirac gluino is shown in fig.~\ref{fig:dirac}.
For the stop mass of $2$ TeV, $\mathcal{O}(10)$\% tuning is possible even if the mediation scale is as high as $10^{16}$ GeV.
Note that in Dirac gluino models the large log enhancement of the quantum correction to the Higgs mass squared is already absent.
Thus the improvement of the fine-tuning by the Twin higgs mechanism simply originates from a large $SU(4)$ invariant coupling. For $g_X =1 - 1.5$, the improvement is by a factor of $2-4$.

In some UV completions of the Dirac gluino, the fine-tuning may be worse and at the $O(1)\%$ level~\cite{Arvanitaki:2013yja}.
For example in gauge mediated models, a tachyonic soft mass term of the adjoint chiral superfield larger than the Dirac gluino mass is often generated. See ref.~\cite{Csaki:2013fla} for a pedagogical discussion. 
To prevent the instability of the adjoint field one needs to cancel the tachyonic mass by additional large soft mass or a supersymmetric mass of the adjoint, which leads to fine-tuning. See ref.~\cite{Alves:2015kia} for a gauge mediated model free from this problem.
In gravity mediated model the tachyonic mass is not necessarily larger than the Dirac gluino mass.
Our D-term model, together with the Dirac gluino, realizes the natural SUSY even for the gravity mediation.

\begin{figure}[t]
\centering
\includegraphics[clip,width=.48\textwidth]{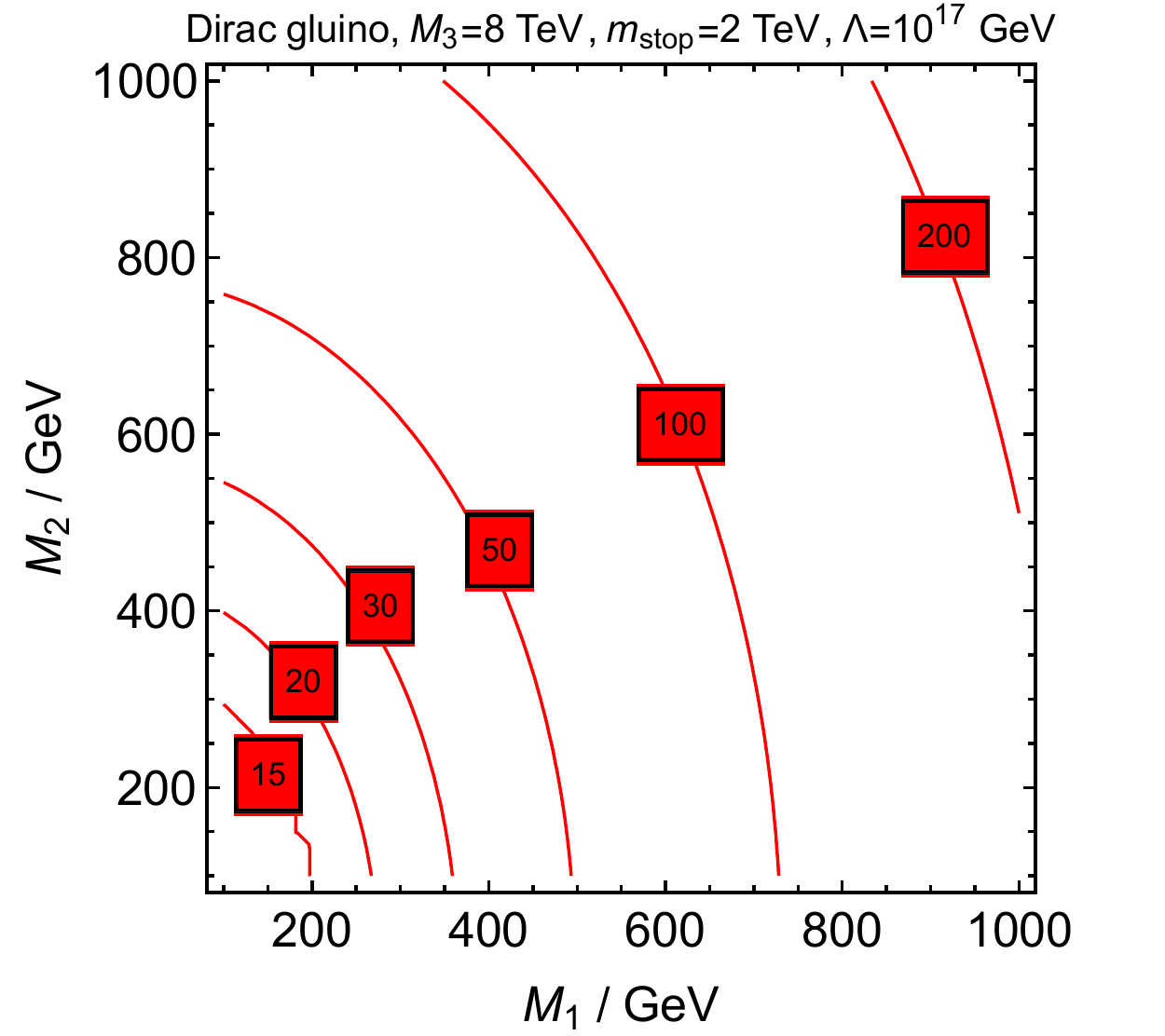}
\includegraphics[clip,width=.48\textwidth]{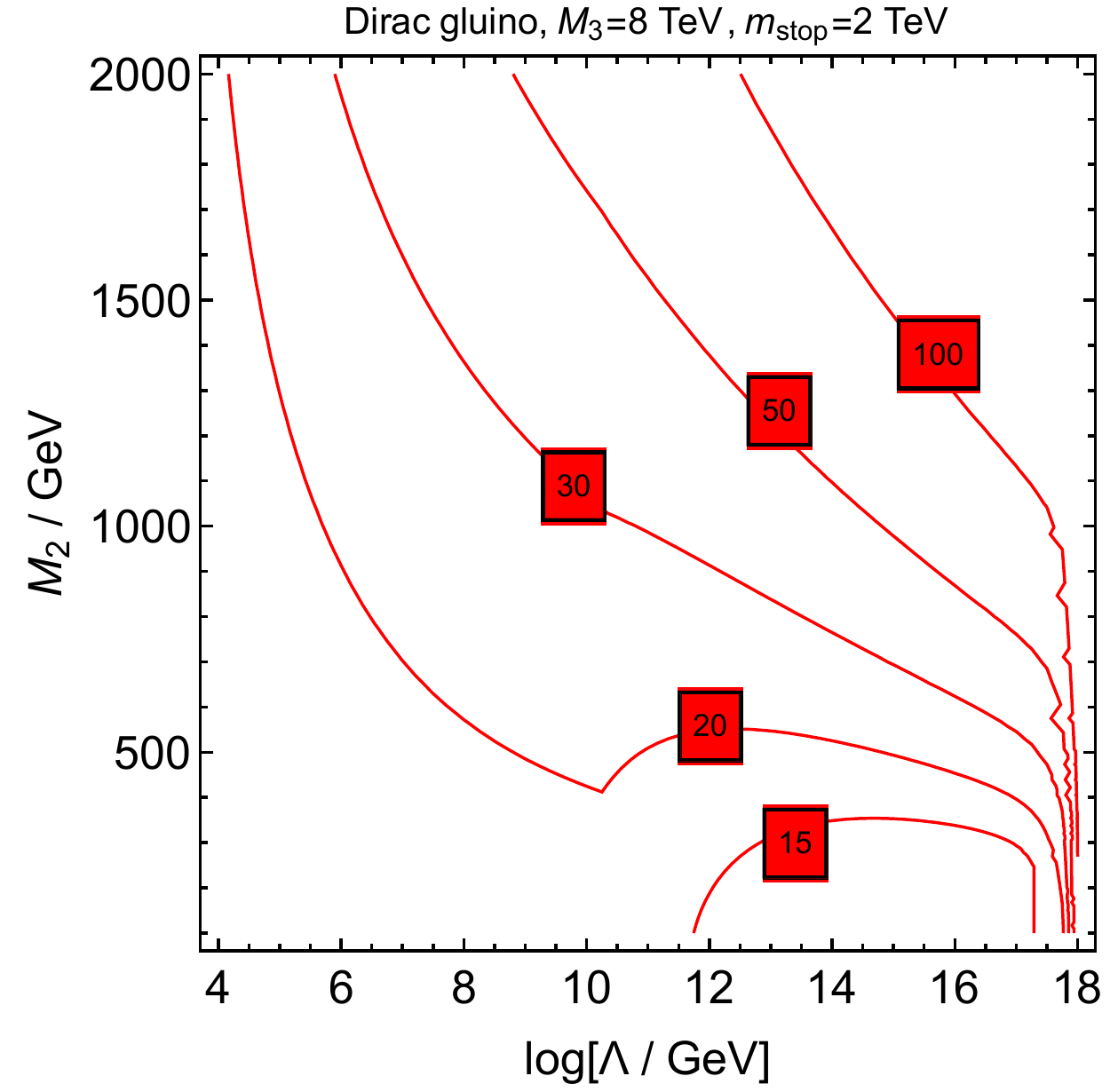}
\caption{
Fine-tuning in the plane $M_1-M_2$ with a Dirac gluino for $\Lambda=10^{17}$ GeV and maximal value of the $SU(2)_X$ gauge coupling constant,
$g_X^{\rm max}$, that do not induce the Landau pole below the mediation scale (left panel). In the right panel, fine-tuning in the plane $\Lambda-M_2$
is shown for $M_1=100$ GeV and $g_X={\rm min}(1.5,g_X^{\rm max})$. The remaining parameters are the same as in fig.~\ref{fig:dirac}. 
}
\label{fig:M1M2}
\end{figure}

The wino and the bino masses are also bounded from above by naturalness. The constraint is stronger than that in the MSSM as we add extra $SU(2)_L$
and/or $U(1)_Y$
charged fields which makes the corresponding gauge couplings and gaugino masses growing faster with the renormalization scale. Fine-tuning from bino
and wino may be very large especially for high mediation scales. In the left panel of fig.~\ref{fig:M1M2} we fix $\Lambda=10^{17}$ GeV and present
contours of fine-tuning in the plane $M_1-M_2$. We see that bino as light as 700~GeV induces tuning at the level of 1~$\%$ for this mediation scale.
The tuning from wino is slightly smaller but still 1~TeV wino results in about 1~$\%$ tuning. From the comparison of
figs.~\ref{fig:dirac}~and~\ref{fig:M1M2} we see that in order not to increase tuning by more than a factor of two, bino (wino) must be lighter than
about 400 (600) GeV. Thus, one generally expects all neutralinos to be light and the LSP to be a mixture of bino, wino and
higgsino. Assuming majorana gluino, the impact of wino and bino on the tuning is less pronounce but in order not to increase tuning by more than a
factor of two their masses are still expected to be below about 1~TeV, cf.~figs.~\ref{fig:gxLambda}~and~\ref{fig:M1M2}. For smaller mediation scales
the tuning from bino and wino is milder. The tuning from bino is subdominant unless $\Lambda\gtrsim10^{16}$~GeV. In the right panel of
fig.~\ref{fig:M1M2} we present tuning from wino as a function of the mediation scale. We see that even for small mediation scale wino mass should
generally be below 1~TeV in order not to dominate tuning.  
The bounds on the masses is avoided if the wino and the bino also obtain Dirac masses. 
Interestingly, with an additional $SU(2)_L$ adjoint paired with wino, the $SU(2)_L$ gauge coupling constant also blows up around the Planck
scale.

\begin{figure}[t]
\centering
\includegraphics[clip,width=.48\textwidth]{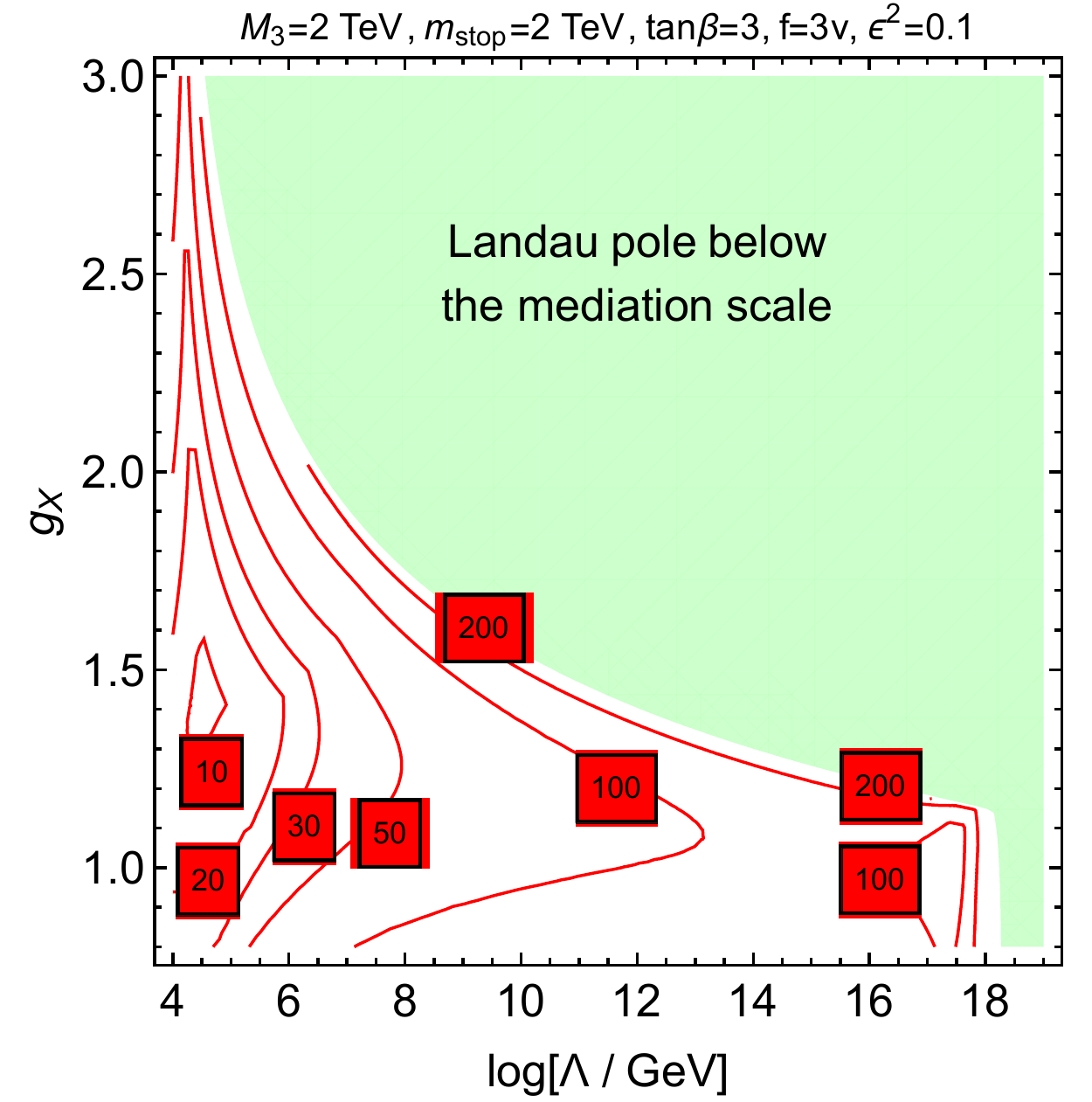}
\includegraphics[clip,width=.48\textwidth]{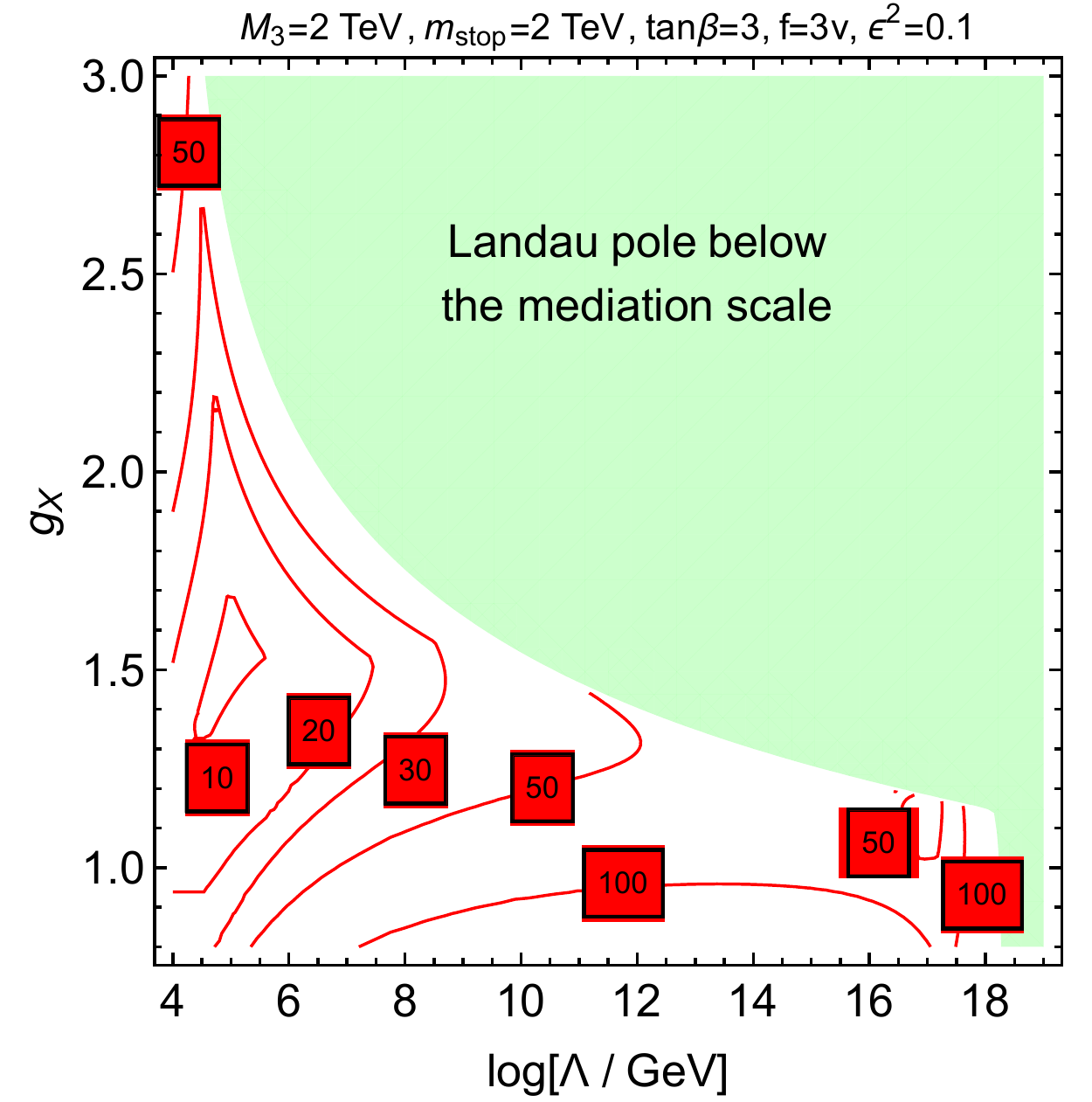}
\caption{
The same as in the left panel of fig.~\ref{fig:gxLambda} but including the effect of $m_S^2$ under the assumption that it is constant during the RGE
flow (left panel) or assuming that $m_S^2=0$ at the mediation scale and large value of $m_S^2$ at low energy corresponding to $\epsilon^2=0.1$ is
obtained due to RG contribution from the soft mass
$m_Z^2$ for $\kappa=0.3$ (right panel).
}
\label{fig:mSeffect}
\end{figure}

In the above analysis we have ignored the contribution to the RGE running of $m_{H_u}^2$ proportional to the $SU(2)_X$ gauge coupling constant.
As long as the $SU(2)_X$ gaugino mass is suppressed, one-loop contributions are negligible. At the two loop level, there is a contribution,
\begin{align}
\frac{{\rm d}}{{\rm dln} \mu} m_{H_u}^2 \supset \frac{3 g_X^4}{256\pi^4} \sum_{i}m_i^2,
\end{align}
where $m_i^2$ is a soft mass squared of a $SU(2)_X$ fundamental.
Although this is a two-loop effect, the largeness of $m_S^2$ required to obtain a large non-decoupling $SU(4)$ invariant quartic and the largeness of 
$g_X$ around the Landau pole scale can make this 
contribution non-negligible.
In the left panel of fig.~\ref{fig:mSeffect}, we show the fine-tuning including this two-loop effect to $m_{H_u}^2$, with $m_S^2$ fixed at the value
determined by eq.~(\ref{eq:epsilon}), while ignoring contribution from other $SU(2)_X$
charged fields.
The fine-tuning gets worse than the case ignoring the two-loop effect, especially when the mediation scale is close to the Landau pole scale, while it
remains the same if the mediation scale 
is much smaller than the Landau pole scale.
We note, however, that the two-loop effect strongly depends on the boundary condition of soft masses at the mediation scale and might be much smaller.
For example, if $m_S^2 = - m_{\bar{E}}^2$ at the boundary, the two-loop effect is suppressed.
This special boundary condition should be explained by a UV completion of our model.
It is also possible that $m_S^2$ at the UV boundary is much smaller than around the $m_X$ scale, and is generated through the RG
running.
Actually couplings with the fields $Z$ and/or $E_{1,2}$ can generate a non-zero and positive $m_S^2$, if the soft masses of them are negative. For
example, in the right panel of fig.~\ref{fig:mSeffect} we show tuning under the assumption that $m_S^2$ vanishes at the mediation scale and gets
renormalized to appropriately large value determined by eq.~(\ref{eq:epsilon}) at the EW scale via the interaction~\eqref{eq:WkappaZSS} (by suitable
choice of the soft mass for $Z$) with $\kappa =0.3$ at the UV boundary. In this case the impact of $m_S^2$ on tuning is rather small unless the mediation scale is very close to the Landau pole scale.

Even though naturalness does not require sparticles to be within discovery reach of the LHC (perhaps except for wino if the mediation scale is high
enough) it does require that the twin Higgs boson is relatively
light.
The mass of the twin Higgs boson is well approximated by $2\sqrt{\lambda}f$ with $\lambda$ being the $SU(4)$ invariant coupling. Both $\lambda$ and $f$
are constrained from above by naturalness. For example, demanding better than $10\%$ tuning $f$ must be below about 4.5. This upper bound on $f$ is
quite generic for Twin Higgs models unless hard $\mathbb{Z}_2$ breaking is non-negligible. The upper bound on $\lambda$ is specific for this model and
is set by the requirement of not too large threshold correction from $X$ gauge bosons. We find that better  than $10\%$ tuning requires
$\lambda\lesssim0.5$ which, together with the upper bound on $f$, leads to the upper bound on the twin Higgs boson mass of about 1 TeV. The twin Higgs
tends to be lighter for a larger Landau pole scale.  For recent studies of the phenomenology  of the twin Higgs boson we refer the 
reader to refs.~\cite{Buttazzo:2015bka,Katz:2016wtw}. It is also noteworthy that in this model MSSM-like Higgs bosons and their mirror counterparts are not required to be light by naturalness because $H_d$ is not charged under $SU(2)_X$.

\section{Summary}

We proposed a new SUSY Twin Higgs model in which an $SU(4)$ invariant quartic term originates from a $D$-term potential of a new $SU(2)_X$ gauge symmetry. The choice of the non-abelian gauge symmetry, together with a minimal number of flavors charged under $SU(2)_X$, makes the running of the new gauge coupling constant rather slow allowing for a large $SU(4)$ invariant quartic term without generating a low-scale Landau pole. The Twin Higgs mechanism, together with the negative contribution from the new gauge coupling to the RG running of the top yukawa coupling, allows for tuning of the EW scale better than $10\%$ for high mediation scales up to $\mathcal{O}(10^9)$ GeV even for sparticle spectra that may be outside of the ultimate LHC reach.
If the gluino obtains a Dirac mass term, tuning of $10\%$ is possible even if  the mediation scale is around the Planck scale.
The model may be tested at the LHC by searching for a twin Higgs boson whose mass is bounded from above by naturalness and is anti-correlated with the
Landau Pole scale.
In parts of parameter space with tuning better than $10\%$ the twin Higgs boson is expected to be lighter than about 1 TeV. All electroweakinos are
expected to be rather light, with masses in the sub-TeV region, especially if the mediation scale of SUSY breaking is high. 

\section*{Acknowledgments}
This work has been partially supported by National Science Centre, Poland, under research grant DEC-2014/15/B/ST2/02157, by
the Office of High Energy Physics of the U.S. Department of Energy
under Contract DE-AC02-05CH11231, and by the National Science Foundation
under grant PHY-1316783. MB acknowledges support from the
Polish 
Ministry of Science and Higher Education through its programme Mobility Plus (decision no.\ 1266/MOB/IV/2015/0). 
The work of KH was in part performed at the Aspen Center for Physics, which is supported by National Science Foundation grant PHY-1607611.

\appendix

\section{Electroweak precision measurements}
\label{sec:EWPM}

\begin{figure}[tb]
\centering
\includegraphics[clip,width=.48\textwidth]{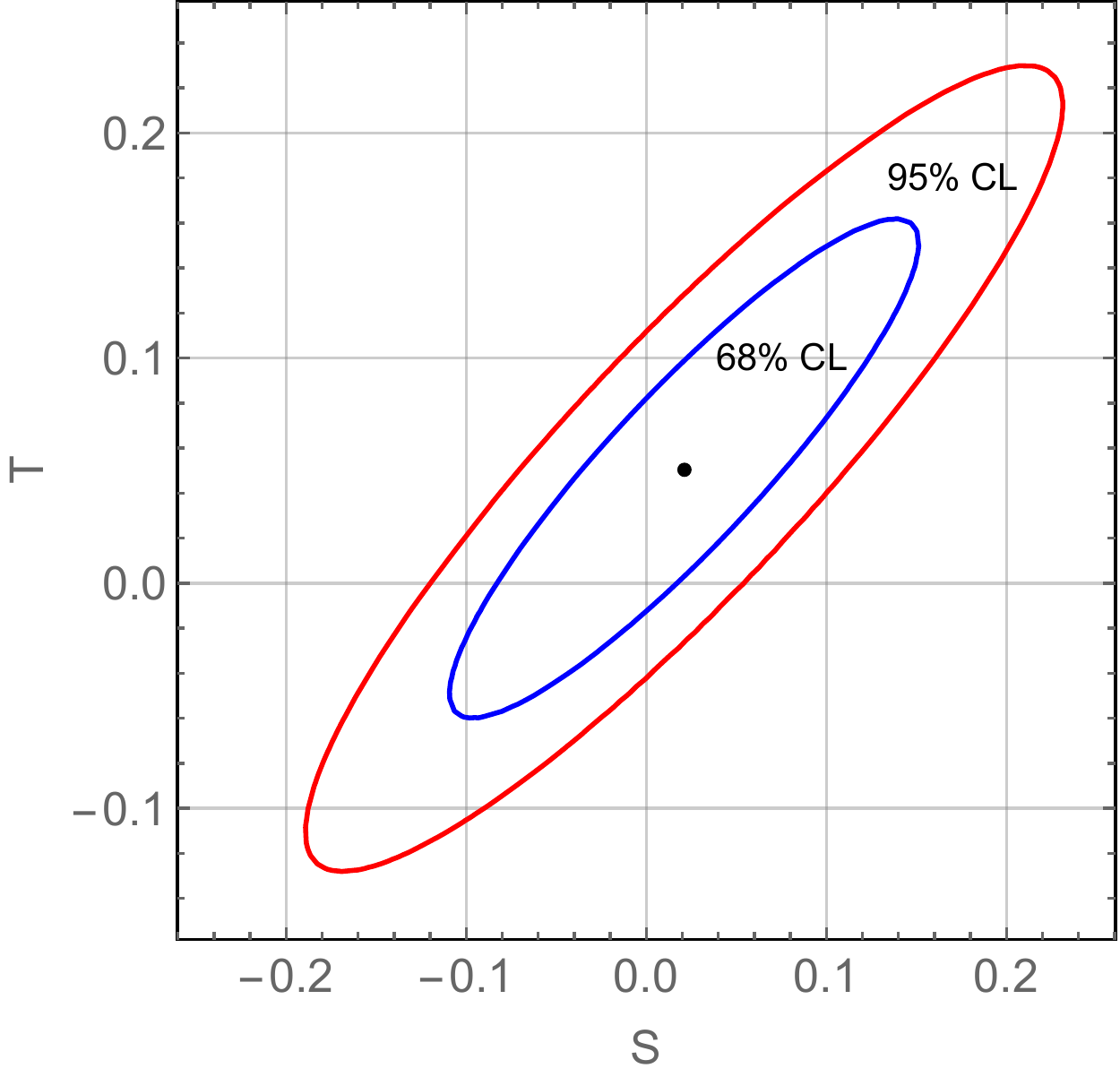}
\caption{The constraint on the S,T parameters.
}
\label{fig:ST}
\end{figure}

We use the so-called $S,T,U$ parametrization~\cite{Peskin:1991sw} to constrain the parameter space of our model.
We follow the method presented in~\cite{Peskin:2001rw}, where the observables shown in Table~\ref{tab:observable} are used to constrain $S,T,U$.
We take $U=0$ and show the constraint on $(S,T)$ in Fig.~\ref{fig:ST}.

The Higgs multiplet ${\cal H}$ is charged under $SU(2)_X$. After the electroweak symmetry breaking scale, $Z$ boson mixes with the $SU(2)_X$ gauge bosons.
The mixing breaks the custodial symmetry and we expect a severe constraint from the electroweak precision measurement.
After integrating out the $SU(2)_X$ gauge bosons, we obtain the effective dimension 6 operator,
\begin{align}
{\cal L}_{\rm eff} = \frac{g_X^2}{8 m_X^2}\left( H_u^\dag D_\mu H_u - (D_\mu H_u)^\dag H_u \right)^2.
\end{align}
This generates a non-zero $T$ parameter,
\begin{align}
T  =  \frac{1}{2\alpha} \frac{c_W^2}{g_2^2} \frac{g_X^2 m_Z^2}{ m_X^2} \sin^2\beta\,.
\end{align}
The dependence on $\tan\beta$ originates from subdominant $H_d$ component of the Higgs.  
$S$ and $U$ parameters are negligibly small. Comparing this result with fig.~\ref{fig:ST}, we obtain the constraint,
\begin{align}
m_X / g_X > 4.1~{\rm TeV} \sin^2\beta.
\end{align}

\begin{table}[htp]
\caption{Values of observable use to constraint the $STU$ parameters.}
\begin{center}
\begin{tabular}{|c|c|c|}
\hline 
observables & value & reference \\ \hline
$\alpha^{-1}$ & $137.035 999 139(31)$ & \cite{Mohr:2015ccw} \\
$G_F({\rm GeV}^{-2})$ & $1.1663787(6)\times 10^{-5}$ &  \cite{Mohr:2015ccw} \\
$m_Z$ (GeV) & $91.1876(21)$ & \cite{Olive:2016xmw} \\
$\Delta \alpha_{\rm lep}(m_Z^2)$ & $0.03150$ & \cite{Steinhauser:1998rq} \\
$\Delta \alpha_{\rm had}(m_Z^2)$ & $0.02764(13)$ & \cite{Olive:2016xmw} \\
$m_W$ (GeV) & $80.385(15)$ &  \cite{Olive:2016xmw} \\
$m_t$ (GeV) & $173.3 (8)$ & \cite{ATLAS:2014wva} \\
$m_h$ (GeV) & $125$ & \cite{Higgsmass_exp}  \\
$\bar{s}_\ell^2$ & $0.23153(16)$ & \cite{ALEPH:2005ab}  \\
$\Gamma_{Z\rightarrow \ell^+\ell^-}$ (MeV) & $83.984(86)$  & \cite{Olive:2016xmw} \\
$\alpha_s(m_Z^2)$ & $0.1181(11)$ &  \cite{Olive:2016xmw} \\ \hline
\end{tabular}
\end{center}
\label{tab:observable}
\end{table}%

The mixing between the SM-like Higgs and the mirror Higgs also contributes to  $S$ and $T$ parameters,
\begin{align}
S =  \frac{1}{12\pi} s_\gamma^2 {\rm ln} \frac{m_{h'}^2}{m_h^2},\\
T = - \frac{3}{16\pi c_w^2} s_\gamma^2 {\rm ln} \frac{m_{h'}^2}{m_h^2}.
\end{align}
where $\gamma $ is the mixing angle between the SM Higgs and the mirror Higgs.
Note that the sign of the T parameter is negative, and hence this contribution relaxes the constraint on the $SU(2)_X$ symmetry breaking scale.
In Sec.~\ref{sec:FT}, we use the constraint including the Higgs mixing to determine the magnitude of the threshold correction around the $SU(2)_X$ symmetry breaking scale.


\begin{thebibliography}{99}

\bibitem{MaianiLecture}
L.~Maiani. in Proceedings: Summer School on Particle Physics, Paris,
	France (1979).
\bibitem{Veltman:1980mj} 
  M.~J.~G.~Veltman,
  Acta Phys.\ Polon.\ B {\bf 12}, 437 (1981).
\bibitem{Witten:1981nf} 
  E.~Witten,
  Nucl.\ Phys.\ B {\bf 188}, 513 (1981).
\bibitem{Kaul:1981wp} 
  R.~K.~Kaul,
  Phys.\ Lett.\ B {\bf 109}, 19 (1982).


\bibitem{Buckley:2016kvr}
  M.~R.~Buckley, D.~Feld, S.~Macaluso, A.~Monteux and D.~Shih,
  arXiv:1610.08059 [hep-ph].

\bibitem{Buckley:2016tbs}
  M.~R.~Buckley, A.~Monteux and D.~Shih,
  JHEP {\bf 1706} (2017) 103
  [arXiv:1611.05873 [hep-ph]].


\bibitem{Dimopoulos:2014aua}
  S.~Dimopoulos, K.~Howe and J.~March-Russell,
  Phys.\ Rev.\ Lett.\  {\bf 113} (2014) 111802
  [arXiv:1404.7554 [hep-ph]].
  
  \bibitem{Cohen:2015ala}
  T.~Cohen, J.~Kearney and M.~Luty,
  Phys.\ Rev.\ D {\bf 91} (2015) 075004
  [arXiv:1501.01962 [hep-ph]].
  
\bibitem{Martin:2015eca}
  S.~P.~Martin,
  Phys.\ Rev.\ D {\bf 92} (2015) no.3,  035004
  [arXiv:1506.02105 [hep-ph]].

\bibitem{Garcia:2015sfa}
  I.~Garcia Garcia, K.~Howe and J.~March-Russell,
  JHEP {\bf 1512} (2015) 005
  [arXiv:1510.07045 [hep-ph]].
  
\bibitem{Chacko:2005pe}
  Z.~Chacko, H.~S.~Goh and R.~Harnik,
  Phys.\ Rev.\ Lett.\  {\bf 96} (2006) 231802
  [hep-ph/0506256].

\bibitem{Chacko:2005vw} 
  Z.~Chacko, Y.~Nomura, M.~Papucci and G.~Perez,
  JHEP {\bf 0601}, 126 (2006)
  [hep-ph/0510273].

\bibitem{Chacko:2005un} 
  Z.~Chacko, H.~S.~Goh and R.~Harnik,
  JHEP {\bf 0601}, 108 (2006)
  [hep-ph/0512088].

\bibitem{Falkowski:2006qq}
  A.~Falkowski, S.~Pokorski and M.~Schmaltz,
  Phys.\ Rev.\ D {\bf 74} (2006) 035003
  [hep-ph/0604066].


\bibitem{Chang:2006ra}
  S.~Chang, L.~J.~Hall and N.~Weiner,
  Phys.\ Rev.\ D {\bf 75} (2007) 035009
  [hep-ph/0604076].
    
 
\bibitem{Batra:2008jy}
  P.~Batra and Z.~Chacko,
  Phys.\ Rev.\ D {\bf 79} (2009) 095012
  [arXiv:0811.0394 [hep-ph]].

\bibitem{Geller:2014kta} 
  M.~Geller and O.~Telem,
  Phys.\ Rev.\ Lett.\  {\bf 114}, 191801 (2015)
  [arXiv:1411.2974 [hep-ph]].
  
  \bibitem{Barbieri:2015lqa}
  R.~Barbieri, D.~Greco, R.~Rattazzi and A.~Wulzer,
  JHEP {\bf 1508} (2015) 161
  [arXiv:1501.07803 [hep-ph]].
  
  \bibitem{Low:2015nqa}
  M.~Low, A.~Tesi and L.~T.~Wang,
  Phys.\ Rev.\ D {\bf 91} (2015) 095012
  [arXiv:1501.07890 [hep-ph]].

\bibitem{Cheng:2015buv} 
  H.~C.~Cheng, S.~Jung, E.~Salvioni and Y.~Tsai,
  JHEP {\bf 1603}, 074 (2016)
  [arXiv:1512.02647 [hep-ph]].

\bibitem{Csaki:2015gfd} 
  C.~Csaki, M.~Geller, O.~Telem and A.~Weiler,
  JHEP {\bf 1609}, 146 (2016)
  [arXiv:1512.03427 [hep-ph]].

\bibitem{Cheng:2016uqk} 
  H.~C.~Cheng, E.~Salvioni and Y.~Tsai,
  Phys.\ Rev.\ D {\bf 95}, no. 11, 115035 (2017)
  [arXiv:1612.03176 [hep-ph]].

\bibitem{Contino:2017moj} 
  R.~Contino, D.~Greco, R.~Mahbubani, R.~Rattazzi and R.~Torre,
  arXiv:1702.00797 [hep-ph].


\bibitem{Barbieri:2016zxn} 
  R.~Barbieri, L.~J.~Hall and K.~Harigaya,
  JHEP {\bf 1611}, 172 (2016)
  [arXiv:1609.05589 [hep-ph]].
 
\bibitem{Chacko:2016hvu}
  Z.~Chacko, N.~Craig, P.~J.~Fox and R.~Harnik,
  [arXiv:1611.07975 [hep-ph]].

\bibitem{Craig:2016lyx} 
  N.~Craig, S.~Koren and T.~Trott,
  JHEP {\bf 1705}, 038 (2017)
  [arXiv:1611.07977 [hep-ph]].

\bibitem{Garcia:2015loa} 
  I.~Garcia Garcia, R.~Lasenby and J.~March-Russell,
  Phys.\ Rev.\ D {\bf 92}, no. 5, 055034 (2015)
  [arXiv:1505.07109 [hep-ph]].

\bibitem{Garcia:2015toa} 
  I.~Garcia Garcia, R.~Lasenby and J.~March-Russell,
  Phys.\ Rev.\ Lett.\  {\bf 115}, no. 12, 121801 (2015)
  [arXiv:1505.07410 [hep-ph]].

\bibitem{Craig:2015xla} 
  N.~Craig and A.~Katz,
  JCAP {\bf 1510}, no. 10, 054 (2015)
  [arXiv:1505.07113 [hep-ph]].

\bibitem{Farina:2015uea} 
  M.~Farina,
  JCAP {\bf 1511}, no. 11, 017 (2015)
  [arXiv:1506.03520 [hep-ph]].

\bibitem{Freytsis:2016dgf} 
  M.~Freytsis, S.~Knapen, D.~J.~Robinson and Y.~Tsai,
  JHEP {\bf 1605}, 018 (2016)
  [arXiv:1601.07556 [hep-ph]].

\bibitem{Farina:2016ndq} 
  M.~Farina, A.~Monteux and C.~S.~Shin,
  Phys.\ Rev.\ D {\bf 94}, no. 3, 035017 (2016)
  [arXiv:1604.08211 [hep-ph]].
 
\bibitem{Prilepina:2016rlq}
  V.~Prilepina and Y.~Tsai,
  arXiv:1611.05879 [hep-ph].

\bibitem{Barbieri:2017opf} 
  R.~Barbieri, L.~J.~Hall and K.~Harigaya,
  arXiv:1706.05548 [hep-ph].


\bibitem{Craig:2013fga}
  N.~Craig and K.~Howe,
  JHEP {\bf 1403} (2014) 140
  [arXiv:1312.1341 [hep-ph]].

\bibitem{Katz:2016wtw}
  A.~Katz, A.~Mariotti, S.~Pokorski, D.~Redigolo and R.~Ziegler,
  JHEP {\bf 1701} (2017) 142
  [arXiv:1611.08615 [hep-ph]].

\bibitem{Badziak:2017syq}
  M.~Badziak and K.~Harigaya,
  JHEP {\bf 1706} (2017) 065
  [arXiv:1703.02122 [hep-ph]].

\bibitem{Higgscomb}
  G.~Aad {\it et al.} [ATLAS and CMS Collaborations],
  JHEP {\bf 1608} (2016) 045
  [arXiv:1606.02266 [hep-ex]].

\bibitem{Buttazzo:2015bka}
  D.~Buttazzo, F.~Sala and A.~Tesi,
  JHEP {\bf 1511} (2015) 158
  [arXiv:1505.05488 [hep-ph]].

\bibitem{Higgsmass_exp}
  G.~Aad {\it et al.} [ATLAS and CMS Collaborations],
  Phys.\ Rev.\ Lett.\  {\bf 114} (2015) 191803
  doi:10.1103/PhysRevLett.114.191803
  [arXiv:1503.07589 [hep-ex]].


\bibitem{Pierce:1996zz}
  D.~M.~Pierce, J.~A.~Bagger, K.~T.~Matchev and R.~j.~Zhang,
  Nucl.\ Phys.\ B {\bf 491} (1997) 3
  [hep-ph/9606211].

\bibitem{Sirunyan:2017cwe}
  A.~M.~Sirunyan {\it et al.} [CMS Collaboration],
  arXiv:1704.07781 [hep-ex].
  
\bibitem{ATLAS:2017vjw}
  The ATLAS collaboration [ATLAS Collaboration],
  ATLAS-CONF-2017-021.
  

\bibitem{Kim:2016rsd}
  J.~S.~Kim, K.~Rolbiecki, R.~Ruiz, J.~Tattersall and T.~Weber,
  Phys.\ Rev.\ D {\bf 94} (2016) no.9,  095013
  [arXiv:1606.06738 [hep-ph]].

\bibitem{Fox:2002bu} 
  P.~J.~Fox, A.~E.~Nelson and N.~Weiner,
  JHEP {\bf 0208}, 035 (2002)
  [hep-ph/0206096].

\bibitem{Arvanitaki:2013yja} 
  A.~Arvanitaki, M.~Baryakhtar, X.~Huang, K.~van Tilburg and G.~Villadoro,
  JHEP {\bf 1403}, 022 (2014)
  [arXiv:1309.3568 [hep-ph]].

\bibitem{Csaki:2013fla} 
  C.~Csaki, J.~Goodman, R.~Pavesi and Y.~Shirman,
  Phys.\ Rev.\ D {\bf 89}, no. 5, 055005 (2014)
  [arXiv:1310.4504 [hep-ph]].

\bibitem{Alves:2015kia} 
  D.~S.~M.~Alves, J.~Galloway, M.~McCullough and N.~Weiner,
  Phys.\ Rev.\ Lett.\  {\bf 115}, no. 16, 161801 (2015)
  [arXiv:1502.03819 [hep-ph]].
  
  \bibitem{Barbieri:2005ri}
  R.~Barbieri, T.~Gregoire and L.~J.~Hall,
  hep-ph/0509242.

\bibitem{Peskin:1991sw} 
  M.~E.~Peskin and T.~Takeuchi,
  Phys.\ Rev.\ D {\bf 46}, 381 (1992).

\bibitem{Peskin:2001rw} 
  M.~E.~Peskin and J.~D.~Wells,
  Phys.\ Rev.\ D {\bf 64}, 093003 (2001)
  [hep-ph/0101342].

\bibitem{Mohr:2015ccw} 
  P.~J.~Mohr, D.~B.~Newell and B.~N.~Taylor,
  Rev.\ Mod.\ Phys.\  {\bf 88}, no. 3, 035009 (2016)
  [arXiv:1507.07956 [physics.atom-ph]].

\bibitem{Olive:2016xmw} 
  C.~Patrignani {\it et al.} [Particle Data Group],
  Chin.\ Phys.\ C {\bf 40}, no. 10, 100001 (2016).

\bibitem{Steinhauser:1998rq} 
  M.~Steinhauser,
  Phys.\ Lett.\ B {\bf 429}, 158 (1998)
  [hep-ph/9803313].

\bibitem{ATLAS:2014wva} 
  [ATLAS and CDF and CMS and D0 Collaborations],
  arXiv:1403.4427 [hep-ex].

\bibitem{ALEPH:2005ab} 
  S.~Schael {\it et al.} [ALEPH and DELPHI and L3 and OPAL and SLD Collaborations and LEP Electroweak Working Group and SLD Electroweak Group and SLD Heavy Flavour Group],
  Phys.\ Rept.\  {\bf 427}, 257 (2006)
  [hep-ex/0509008].

\end{thebibliography}
\end{document}